\documentclass[12pt,letter]{article}
\usepackage{epsfig}
\usepackage{amssymb,amsmath}
\usepackage{setspace}
\usepackage{subfigure}
\usepackage{graphicx}
\usepackage{color}
\usepackage[colorlinks=true
,urlcolor=blue
,citecolor=blue
,linkcolor=blue
,pagecolor=blue
,linktocpage=true
,pdfproducer=medialab
]{hyperref}
\usepackage{longtable}

\def\bi{\begin{itemize}}
\def\ei{\end{itemize}}

\def\tst{\tilde t}

\def\alt{\lesssim}
\def\agt{\gtrsim}

\newcommand\prd[3]{{\it Phys.\ Rev.\ }{\bf D #1} (#2) #3}

\newcommand\npb[3]{{\it Nucl.\ Phys.\ }{\bf B #1} (#2) #3}
\newcommand\plb[3]{{\it Phys.\ Lett.\ }{\bf B #1} (#2) #3}

\newcommand\jhep[3]{{\it J. High Energy Phys.\ }{\bf #1} (#2) #3}
\newcommand\epjc[3]{{\it Eur.\ Phys.\ J. }{\bf C #1} (#2) #3}
\newcommand\jpcs[3]{{\it J. \ Phys. \ Conf. \ Ser.\ }{\bf #1} (#2) #3}
\def\tst{\tilde t}

\newcommand{\bea}{\begin{eqnarray}}
\newcommand{\eea}{\end{eqnarray}}

\newcommand{\beq}{\begin{equation}}
\newcommand{\eeq}{\end{equation}}

 \makeatletter
\def\alt{\mathrel{\mathpalette\gl@align<}}
\def\agt{\mathrel{\mathpalette\gl@align>}}
\def\gl@align#1#2{\lower.6ex\vbox{\baselineskip\z@skip\lineskip\z@
\ialign{$\m@th#1\hfil##\hfil$\crcr#2\crcr\sim\crcr}}} \makeatother

\begin{document}

\begin{titlepage}
\begin{flushright}
UH-511-1229-14
\end{flushright}
\vspace*{1.0cm}

\begin{center}
\baselineskip 20pt

 {\Large\bf
Effects of  Neutrino Inverse Seesaw  Mechanism on the Sparticle Spectrum
in CMSSM and NUHM2}
\vspace{1cm}

{\large
I. Gogoladze$^{a,}$\footnote{E-mail:ilia@bartol.udel.edu},
B. He$^{a,}$\footnote{ E-mail:hebin@udel.edu},
A. Mustafayev$^{b,}$\footnote{ E-mail:azar@phys.hawaii.edu},
S. Raza$^{c,}$\footnote{ E-mail:shabbar@itp.ac.cn}
and Q. Shafi$^{a,}$\footnote{ E-mail:shafi@bartol.udel.edu}
} \vspace{.5cm}

{\baselineskip 20pt \it $^a$
Bartol Research Institute, Department of Physics and Astronomy,
University of Delaware, Newark, DE 19716, USA \\
}
{ \it $^b$ Department of Physics and Astronomy,
University of Hawaii, Honolulu, HI 96822, USA \\
}
{\it $^c$
State Key Laboratory of Theoretical Physics and Kavli Institute for Theoretical Physics China (KITPC),
Institute of Theoretical Physics, Chinese Academy of Sciences, Beijing 100190, P. R. China
}

\end{center}

\vspace{0.5cm}

\begin{abstract}
\noindent
We study the implications of the inverse seesaw mechanism (ISS) on the sparticle spectrum in the
Constrained Minimal Supersymmetric Standard Model (CMSSM) and Non-Universal Higgs Model (NUHM2).
Employing the maximal value of the Dirac Yukawa coupling involving the up type Higgs doublet provides a 2-3~GeV
enhancement of the lightest CP-even Higgs boson  mass.
This effect permits one to have lighter colored sparticles in the CMSSM and NUHM2 scenarios with LSP neutralino,
which can be tested at LHC14. We present a variety of LHC testable benchmark points
with the desired LSP neutralino dark matter relic abundance.

\vspace*{0.8cm}

\end{abstract}

\end{titlepage}


\section{Introduction}

The recent discovery of the
Standard Model (SM)-like Higgs boson with mass $m_h = 125.5\pm 0.5$~GeV
 by the ATLAS~\cite{atlas:2012gk} and  CMS~\cite{cms:2012gu}  experiments at the Large Hadron Collider (LHC) has sparked detailed examinations of viable regions of the parameter space of low scale supersymmetry. This is largely motivated by the fact that the Minimal
Supersymmetric Standard Model (MSSM)  predicts an upper bound on the mass of the lightest CP-even Higgs boson mass, $m_h \lesssim 135$~GeV~\cite{Carena:2002es}. The Higgs boson mass and the corresponding sparticle spectrum strongly depend on
the soft supersymmetry breaking (SSB) parameters~\cite{Djouadi:2005gj}, which can be tested at the LHC (see, for instance \cite{Gogoladze:2011aa, Ajaib:2012vc,Baer:2011ab}).

In low  scale supersymmetry, a Higgs boson mass of around 125~GeV  requires
 either a relatively large value, ${\cal O} (\mathrm{few}-10)$~TeV,
 for the geometric mean of top squark masses~\cite{Ajaib:2012vc}, or a large
SSB trilinear $A_t$-term, with a  geometric mean of the top squark masses of
around a TeV~\cite{Baer:2011ab}.
The presence of heavy top squarks typically yields a heavy sparticle spectrum in gravity mediated supersymmetry  breaking \cite{msugra}, if universality at $M_{\rm GUT}$ of sfermion masses is assumed.
It is especially hard in this case to achieve colored sparticles lighter than 2.5~TeV.

The current LHC lower bounds on the colored sparticle masses from LHC data are
 $ m_{\tilde{g}} \gtrsim  1.5~{\rm~TeV}~ ({\rm for}~ m_{\tilde{g}}\sim m_{\tilde{q}}),~{\rm and}~
m_{\tilde{g}}\gtrsim 0.9~{\rm~TeV}~({\rm for}~ m_{\tilde{g}}\ll
m_{\tilde{q}})$ ~\cite{Aad:2012fqa,Chatrchyan:2012jx},
and it is expected that the LHC14 can test squarks and gluinos with masses up to 3.5~TeV \cite{cms_lim}. In order to be able to reduce the sparticle masses to more accessible values in models with   universal sfermion and gaugino masses, we require additional contributions from new physics, which preserves gauge coupling unification.

Solar and atmospheric neutrino oscillation experiments have established that at least two neutrino states are massive \cite{GonzalezGarcia:2012sz}.
On the theoretical side  the nature of the  physics responsible for
neutrino masses and flavor properties remains largely unknown and is a subject of extensive
investigations~\cite{Mohapatra:2005wg}.
Since our goal is to lower the sparticle mass spectrum while preserving gauge coupling unification, we utilize in this paper the inverse seesaw mechanism (ISS) for generating the light neutrino masses~\cite{Mohapatra:1986aw}.
Introducing only SM singlet fields allows one to realize the ISS mechanism, and all new fields can be below the
TeV scale.  In addition,  we can have ${\cal O} (1)$ Dirac Yukawa couplings involving the up type Higgs doublet.
It has been shown in Refs.~\cite{Gogoladze:2012jp, Guo:2013sna} that the Dirac Yukawa coupling can impact
the lightest  CP-even Higgs boson mass through radiative corrections and
increase it by 2-3~GeV when the additional new fields are SM singlets.
The ISS mechanism can also  be realized using  $SU(2)_W$  weak triplets~\cite{Gogoladze:2012jp},
and in this case the Higgs mass can be enhanced by more than 10~GeV.

In this paper we restrict ourselves to the case of SM singlet fields
since we do not want to disturb  gauge coupling unification.
An enhancement by 2-3~GeV of the CP-even SM-like Higgs boson
mass, as we  will show,  can yield significant reductions of sparticle masses in the Constrained Minimal Supersymmetric Standard Model (CMSSM)~\cite{msugra} and
Non-Universal Higgs Model with  $m_{H_{u}}^2 \neq m_{H_{d}}^2$ (NUHM2)~\cite{nuhm2}.
Here $m_{H_{u}}^2$ and   $m_{H_{d}}^2$ denote the SSB mass square terms for the up and down type MSSM Higgs doublets respectively.

The outline for the rest of the paper is as follows. In section~\ref{model} we briefly
describe the model including the SSB parameters, the range of values employed
in our scan, and the scanning procedure. The relevant experimental
constraints that we have employed are described in section~\ref{sec:scan}.
The results pertaining to CMSSM, CMSSM-ISS and NUHM2-ISS
are discussed in section~\ref{results}, and our conclusions are summarized in section~\ref{conclusions}.

\section{Inverse Seesaw Mechanism and Higgs Boson Mass }
\label{model}

In order to explain non-zero neutrino masses and mixings by the ISS mechanism~\cite{Mohapatra:1986aw},
we supplement the MSSM field content with three pairs of MSSM singlet chiral superfields  ($N^c_i + N_i$),
$i$ = 1, 2, 3, and a singlet chiral superfield $S$ which develops a vacuum expectation value (VEV) comparable to or less than the electroweak scale. The part of the renormalizable superpotential  involving only the new chiral superfields is given by
\begin{eqnarray}
 W \supset  Y_{N_{ij}} N_i^c H_u L_j + {\lambda_{N_{ij}}} S N_i N_j+ m_{ij} N^c_i N_j.
     \label{eq1}
\end{eqnarray}
 Here $Y_{N_{ij}}$ and $\lambda_{N_{ij}}$ are dimensionless couplings  and $m_{ij}$ is a mass term. A non-zero VEV for the scalar component of $S$ generates the lepton-number-violating  term
 $\mu_s  N_i N_j \equiv{\lambda_{N_{ij}}} <S> N_i N_j$ and, as a result,  Majorana masses for the observed neutrinos
 can  be generated.
The coupling ${\lambda_{N_{ij}}} S N_i N_j$ is preferred over the direct mass term $\mu N_i N_j$,
  with the former yielding the desired mass terms for the $N$ fields with a non-zero $<S>$. A singlet chiral superfield $S$ can make it easier to find extension of the SM gauge group
  with help from a suitable symmetry (see, for instance, Refs.~\cite{Gogoladze:2012jp, Dev:2009aw}),
  and avoid terms which otherwise may spoil the ISS mechanism.

The SSB terms pertaining to the fields $N^c_i$ and $N_i$ are given by
\begin{eqnarray}
        {\cal L}^{\rm soft} \supset  m_{N^c}^2\widetilde{N^c}^\dag\widetilde{N^c}
        +m_N^2\widetilde{N}^\dag\widetilde{N}
+\left[A_\nu^{ij}\widetilde{L}_i\widetilde{N^c}_jH_u
+B^{jk}_{m}\widetilde{N^c}_j\widetilde{N}_k
+B_{\mu_N}^{jk}\widetilde{N}_j\widetilde{N}_k+{\rm h.c.}\right],
\label{soft1}
\end{eqnarray}
where the SSB parameters are prescribed at the TeV SUSY breaking scale. In the ISS case there are regions of the  SSB parameter space for which one of the sneutrinos  can be the
lightest  supersymmetric particle (LSP). The phenomenology of models of this kind has been studied in
Ref.~\cite{Dev:2009aw}. In our present work we assume that the lightest neutralino is the LSP,  and a spectrum of this nature can be realized both in the CMSSM and NUHM2 if we assume that all sfermions,
including the $N^c_i$ and $N_i$ fields, have universal SSB mass terms at $M_{\rm GUT}$.

According to the superpotential  in Eq.~(\ref{eq1}),  after  integrating out the ($N^c_i + N_i$)  fields,
the neutrino mass arises from the effective  dimension six operator (Figure. ~1):
\beq
\frac{LLH_uH_uS}{M_6^2}.
\label{eqq2}
\eeq
We assume here that $M_6\delta_{ij} \equiv m_{ij}$ is larger than the electroweak scale.
Also, in Eq.~(\ref{eqq2}) the family and $SU(2)_W$ gauge indices are omitted.

\begin{figure}
\begin{center}
\includegraphics[width=.7\linewidth]{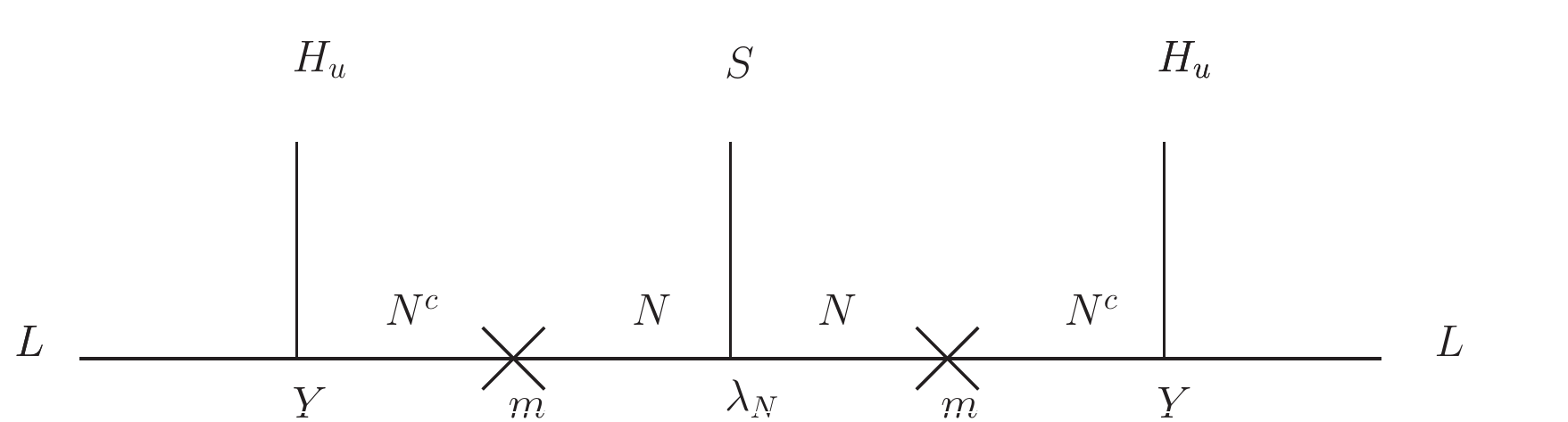}
\label{operator}
\end{center}
\caption{Supergraph leading to dimension six operator for neutrino masses
.}
\end{figure}

 Following the electroweak symmetry breaking, the neutrino Majorana mass matrix is generated:
\beq
  m_\nu = \frac{(Y_N^T Y_N) {\it v_u}^2}{M_6} \times \frac{\lambda_N \langle S \rangle}{M_6}.
  \label{mass2}
\eeq
For simplicity, we set $Y_N\equiv Y_{N_{ij}}$ and $\lambda_N \equiv \lambda_{N_{ij}}$, and $v_u$, $\langle S \rangle$ are the VEVs of
$H_u$, and the $S$ field.
Eq.~(\ref{mass2}) implies that even if we require $Y_N\sim{\cal O}(1)$ and $M_6 \sim 1$~TeV,
the correct mass scale for the  light neutrinos can be reproduced by suitably adjusting
$\lambda_N \langle S \rangle$.

Keeping $Y_N\sim{\cal O}(1)$ will provide sizable contribution to the lightest CP-even Higgs
mass, which is given by~\cite{Babu:2008ge}
\begin{eqnarray}
\left[ m_{h}^{2}\right] _{N} &=&n \times \left[ -M_{Z}^{2}\cos ^{2}2\beta \left(
\frac{1}{8\pi ^{2}} Y_{N} ^{2}t_{N}\right)
+\frac{1}{4\pi ^{2}} Y_{N}^{4}v^{2}\sin ^{4}\beta \left( \frac{1
}{2}\widetilde{X}_{Y_{N}}+t_{N}\right) \right],
\label{e3}
\end{eqnarray}
where
\beq
t_{N}=\log \left( \frac{%
M_{S}^{2}+M_{6}^{2}}{M_{6}^{2}}\right),~
\widetilde{X}_{Y_{N}}=\frac{4\widetilde{A}_{Y_{N}}^{2}\left(
3M_{S}^{2}+2M_{6}^{2}\right) -\widetilde{A}_{Y_{N}
}^{4}-8M_{S}^{2}M_{6}^{2}-10M_{S}^{4}}{6\left(
M_{S}^{2}+M_{6}^{2}\right) ^{2}},
\eeq
and
\beq
\widetilde{A}_{Y_{N}}=A_{Y_{N}}-Y_N \langle S\rangle \cot \beta .
\eeq
Also, $A_{Y_{N}}\equiv A_{\nu}^{ij}$ is the SSB mixing parameter in Eq.~(\ref{soft1}), $n$ is the number of
pairs  of new MSSM singlets, $M_S=\sqrt{m_{\tst_L}m_{\tst_R}}$ defines the SUSY scale, and $v=174.1$~GeV is the electroweak VEV.

We incorporate the ISS mechanism in CMSSM and NUHM2 and scan the SUSY parameter space  using  the ISAJET 7.84
package \cite{isajet}.  We modify  the code by including the additional contributions from  Eq.~(\ref{e3}) to the  lightest CP-even Higgs
boson mass.

\section{Phenomenological constraints and scanning procedure}
\label{sec:scan}

We employ the ISAJET~7.84 package~\cite{isajet}
to generate sparticle spectrum over
the fundamental parameter space.
In this package, the weak scale values of the gauge,  third
 generation Yukawa  couplings,  including  the Yukawa coupling $N^c_i H_u L_j$  from ISS, are evolved to $M_{\rm GUT}$ via the MSSM renormalization group equations (RGEs)
 in the $\overline{DR}$ regularization scheme.
With the boundary conditions given at $M_{\rm GUT}$,
all of the SSB parameters, along with the gauge and Yukawa couplings, are evolved back to the weak scale $M_{\rm Z}$.
The data points collected all satisfy the requirement of radiative electroweak symmetry breaking condition with
 the neutralino in each case being the LSP.

We have performed Markov-chain Monte Carlo (MCMC) scans
 for the following CMSSM parameter range:
\begin{align}
0\leq  m_{0}  \leq 10\, \rm{TeV}, \nonumber \\
0\leq  m_{1/2}  \leq 5\, \rm{TeV}, \nonumber \\
-3\leq A_{0}/m_{0}\leq 3, \nonumber \\
3\leq \tan\beta \leq 60,
 \label{param_range_CMSSM}
\end{align}
with  $\mu > 0$ and  $m_t = 173.3$~GeV \cite{:2009ec}. We use $m_b^{\overline{DR}}(M_{\rm Z})=2.83$~GeV which is hard-coded into ISAJET.  Here $m_0$ is the universal SSB mass parameter for MSSM sfermions, Higgs and additional $N^c$, $N$ and $S$ fields.
$m_{1/2}$ is the gaugino mass parameter,
$\tan\beta$ is the ratio of the VEVs of the two MSSM Higgs
doublets, and $A_0$ is the MSSM universal SSB trilinear scalar coupling. In order to maximize the
contribution from the ISS mechanism to the Higgs boson mass, we set $\tilde{X}_{Y_{N}}=4$, following Ref.~\cite{Gogoladze:2012jp}.

In the case of NUHM2, in addition to the above mentioned parameters we have two additional independent  SSM mass parameters
$m_{H_d}$ and $m_{H_u}$.  We use the following parameter range for them:
\begin{align}
0\leq m_{H_u} \leq 10 \, \rm{TeV}, \nonumber \\
0\leq m_{H_d} \leq 10 \, \rm{TeV}.
\label{param_range_NUHM2}
\end{align}
To maximize the impact of ISS on the sparticle spectrum, we set $\lambda_N=0.7$.
  This is the  maximal value of $\lambda_N$ at low scale that remains perturbative up to $M_{\rm GUT}$.
We also assume that $M_6$ is larger than $M_S$, in order that the neutralino rather than sneutrino is the LSP.

After collecting the data, we impose the mass bounds on all the particles \cite{Beringer:1900zz}  and
 use the IsaTools package~\cite{bsg} and Ref.~\cite{mamoudi}
 to implement the following phenomenological constraints:
\begin{eqnarray}
m_h  = 123-127~{\rm GeV}~~&\cite{ATLAS, CMS}&
\\
0.8\times 10^{-9} \leq{\rm BR}(B_s \rightarrow \mu^+ \mu^-)
  \leq 6.2 \times10^{-9} \;(2\sigma)~~&\cite{Aaij:2012nna}&
\\
2.99 \times 10^{-4} \leq
  {\rm BR}(b \rightarrow s \gamma)
  \leq 3.87 \times 10^{-4} \; (2\sigma)~~&\cite{Amhis:2012bh}&
\\
0.15 \leq \frac{
 {\rm BR}(B_u\rightarrow\tau \nu_{\tau})_{\rm MSSM}}
 {{\rm BR}(B_u\rightarrow \tau \nu_{\tau})_{\rm SM}}
        \leq 2.41 \; (3\sigma)~~&\cite{Asner:2010qj}&.
\end{eqnarray}
As far as the muon anomalous magnetic moment $a_{\mu}$ is concerned, we require that the benchmark
points are at least as consistent with the data as the SM.

For the benchmark points presented in Table 1 and 2, we require that the LSP neutralino dark matter abundance lies in the interval  $0.0913 \leq \Omega_{\rm CDM}h^2  \leq 0.1363$ \cite{WMAP9}.

Finally we implement the following following bounds  on the  sparticle masses:
\begin{align}
 m_{\tilde{g}} \gtrsim  1.5 \, {\rm TeV}\ ({\rm for}\ m_{\tilde{g}}\sim m_{\tilde{q}}) &
 \quad \rm{and}  &
 m_{\tilde{g}}\gtrsim 0.9 \, {\rm TeV}\ ({\rm for}\ m_{\tilde{g}}\ll m_{\tilde{q}}) \quad
\cite{Aad:2012fqa,Chatrchyan:2012jx}.
\end{align}

\begin{figure}[htp!]
\centering
\subfiguretopcaptrue

\subfigure{
\includegraphics[totalheight=5.5cm,width=7.cm]{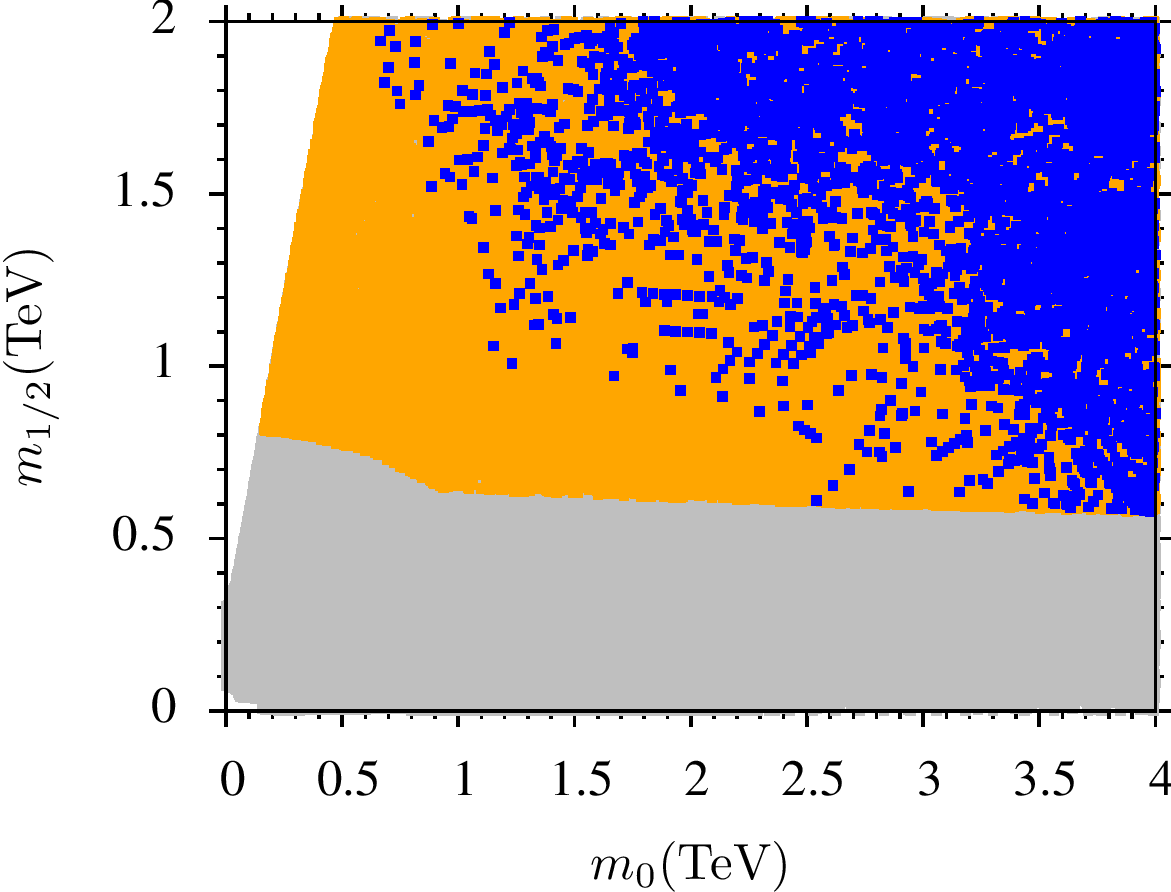}
}
\subfigure{
\includegraphics[totalheight=5.5cm,width=7.cm]{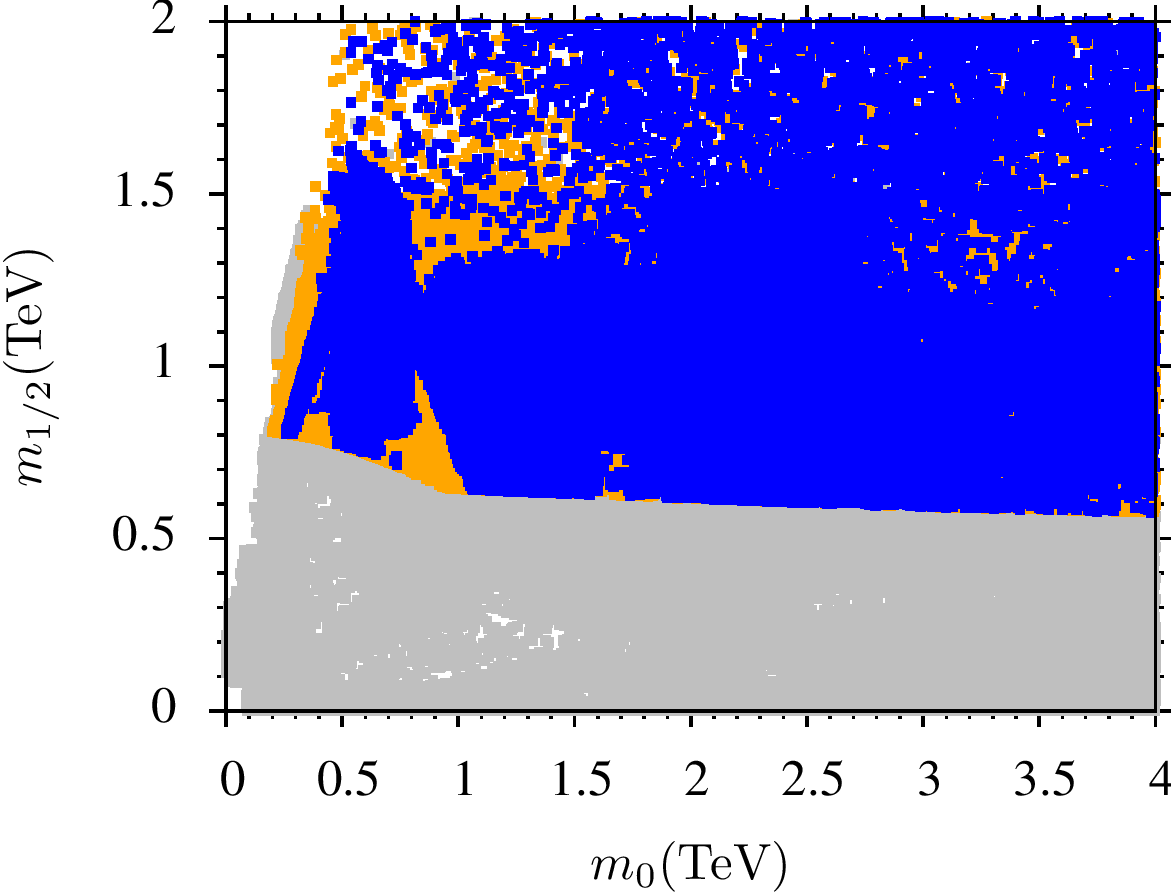}
}
\subfigure{
\includegraphics[totalheight=5.5cm,width=7.cm]{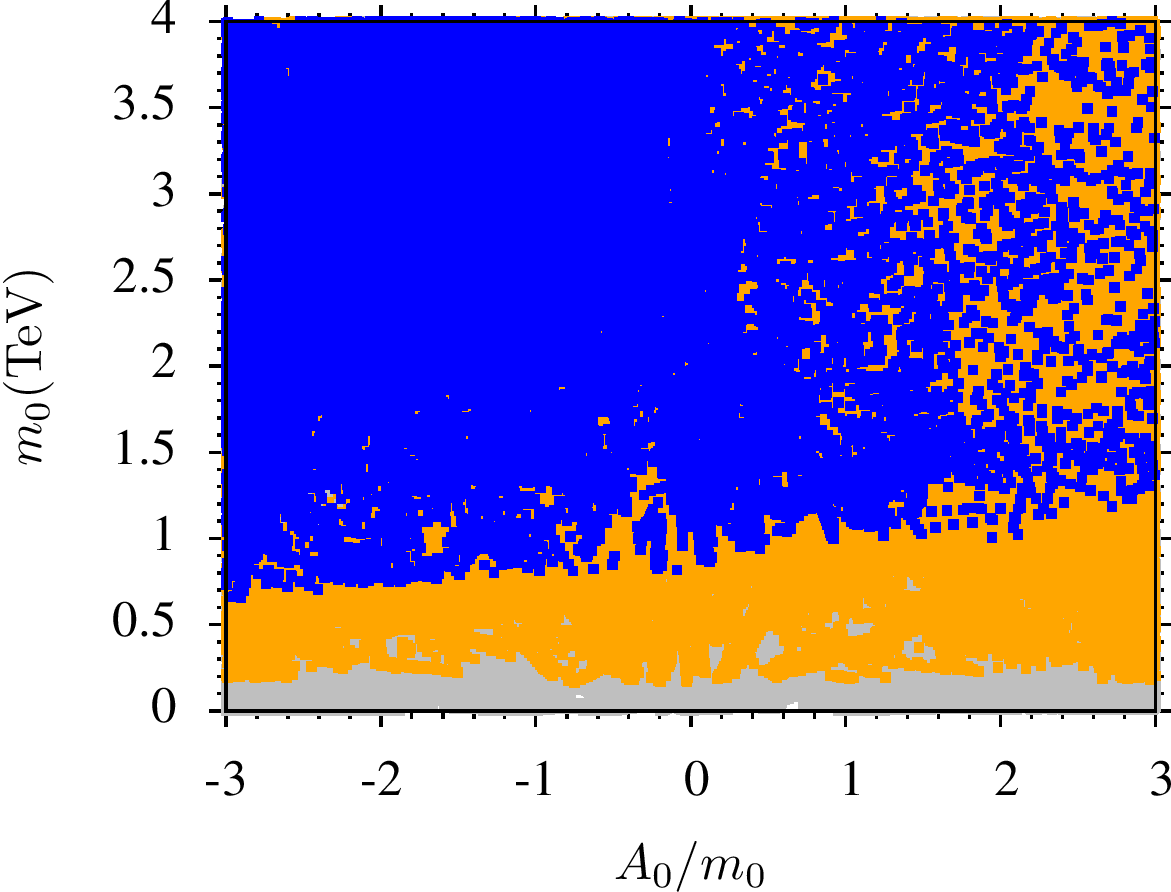}
}
\subfigure{
\includegraphics[totalheight=5.5cm,width=7.cm]{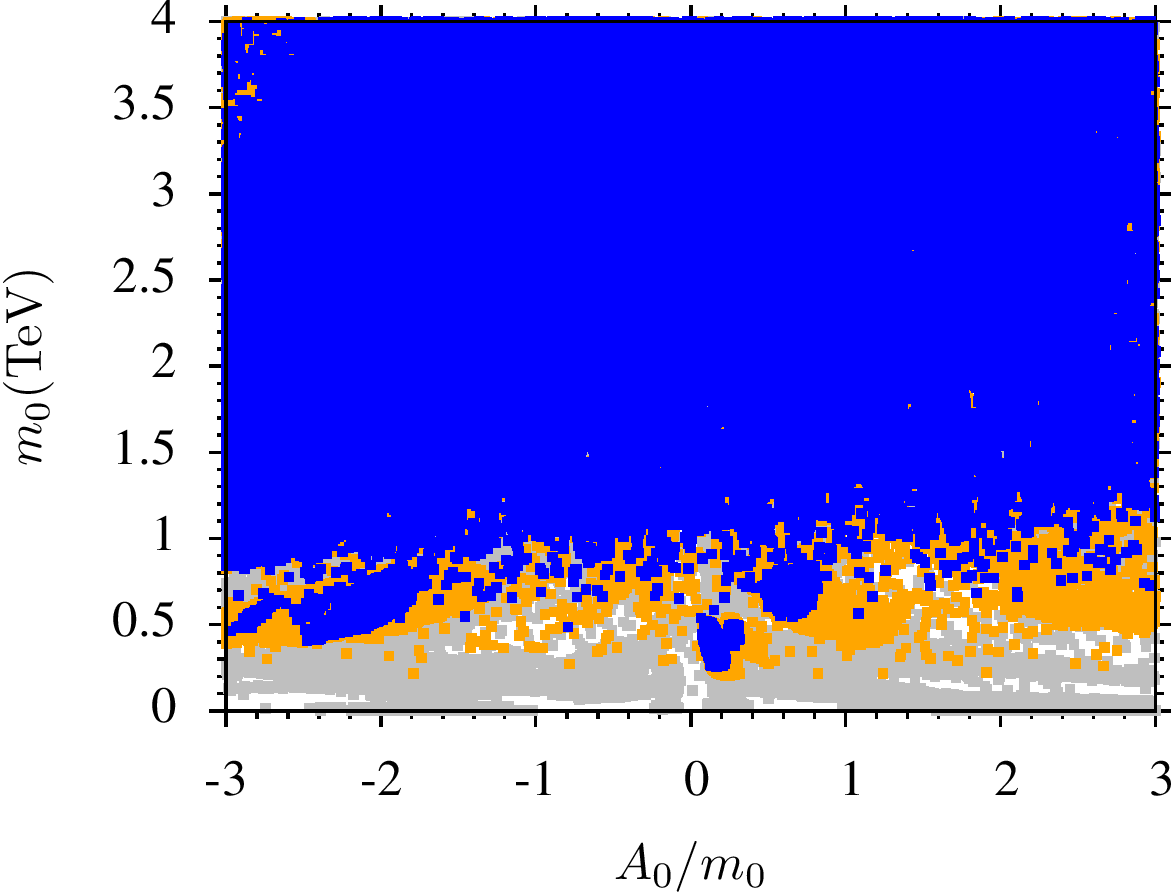}
}
\subfigure{
\includegraphics[totalheight=5.5cm,width=7.cm]{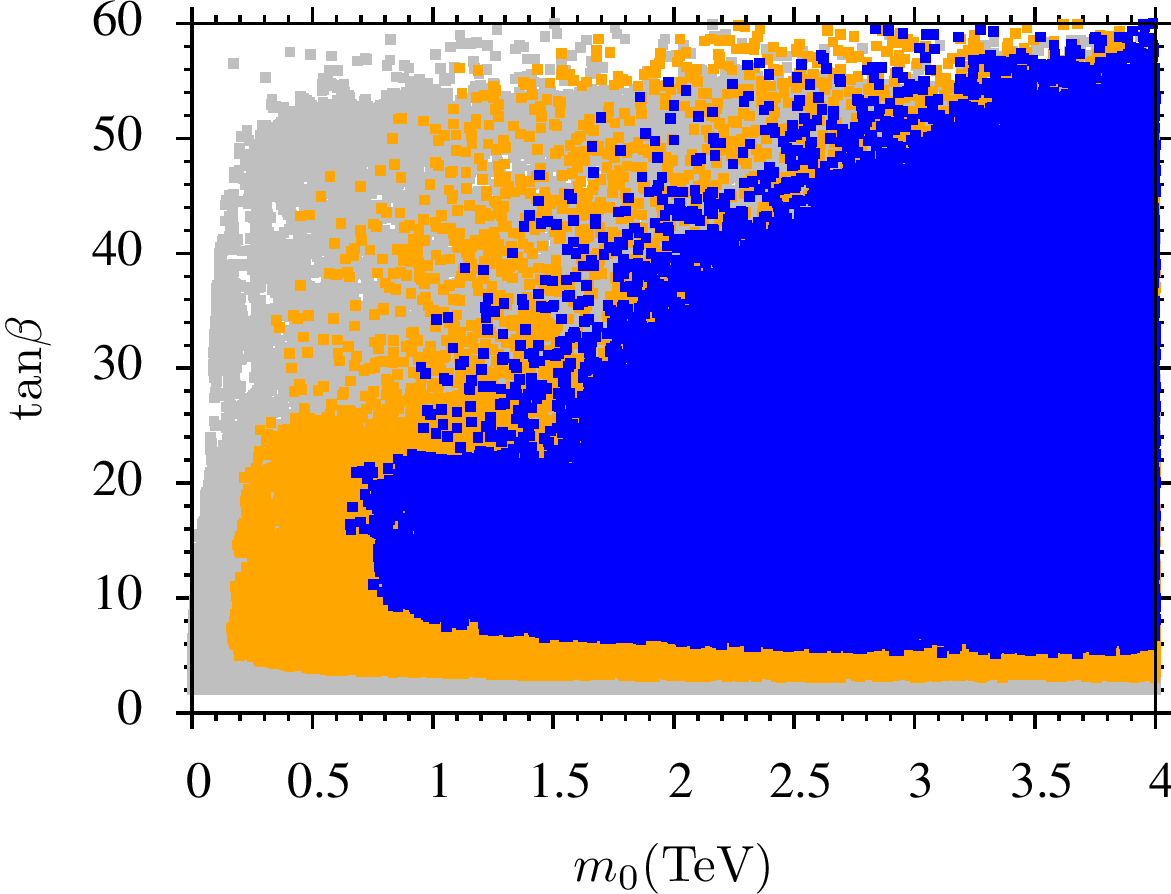}
}
\subfigure{
\includegraphics[totalheight=5.5cm,width=7.cm]{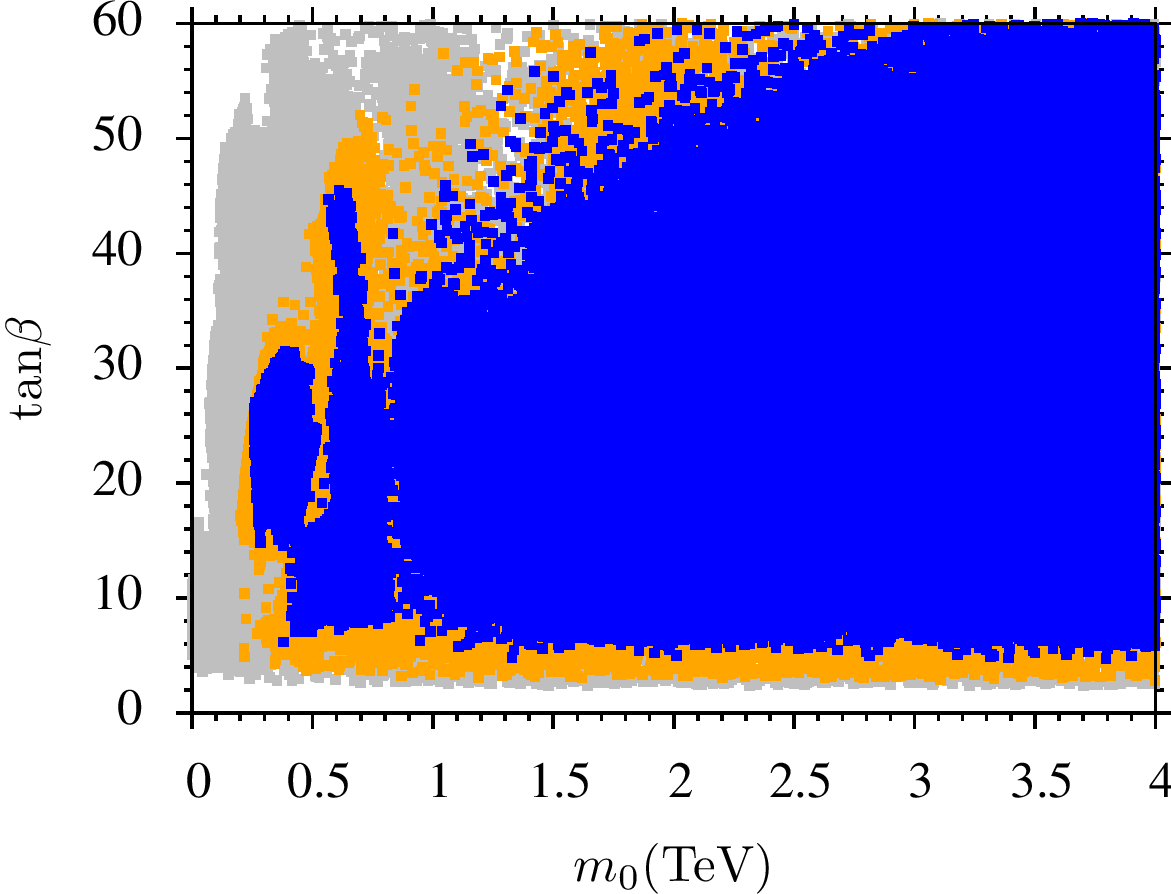}
}

\caption{
Plots in $m_{0}-m_{1/2}$, $A_{0}/m_{0}-m_{0}$ and $m_{0}-\tan\beta$ planes for CMSSM (left panel) and CMSSM-ISS (right panel). Grey points satisfy
REWSB and  LSP neutralino conditions. Orange point solutions satisfy mass bounds and B-physics bounds given in Section 2. Points in blue are a subset of orange points and satisfy $123 \,{\rm GeV}\lesssim m_h\lesssim 127\, {\rm GeV}$.
}
\label{CMSSM_funda}
\end{figure}
\label{sec:results}
\begin{figure}[htp!]
\centering
\subfiguretopcaptrue

\subfigure{
\includegraphics[totalheight=5.5cm,width=7.cm]{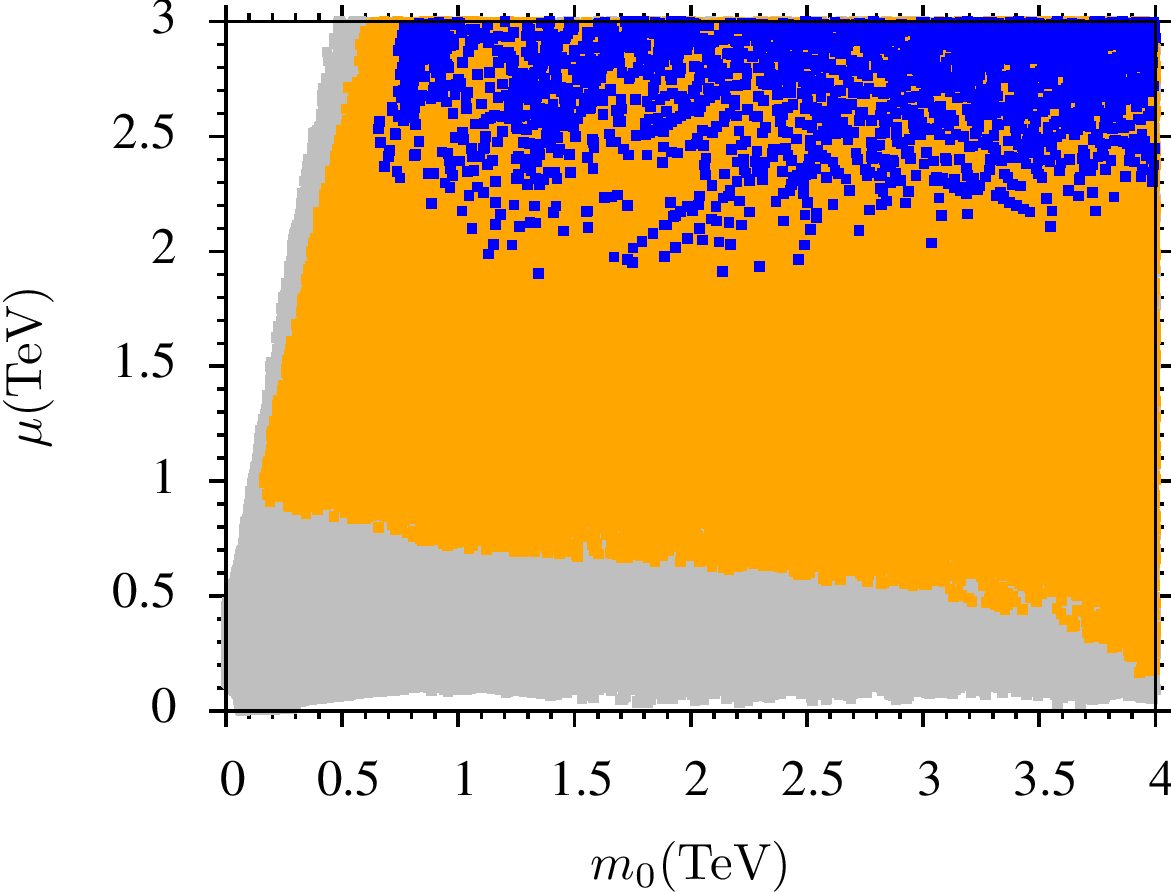}
}
\subfigure{
\includegraphics[totalheight=5.5cm,width=7.cm]{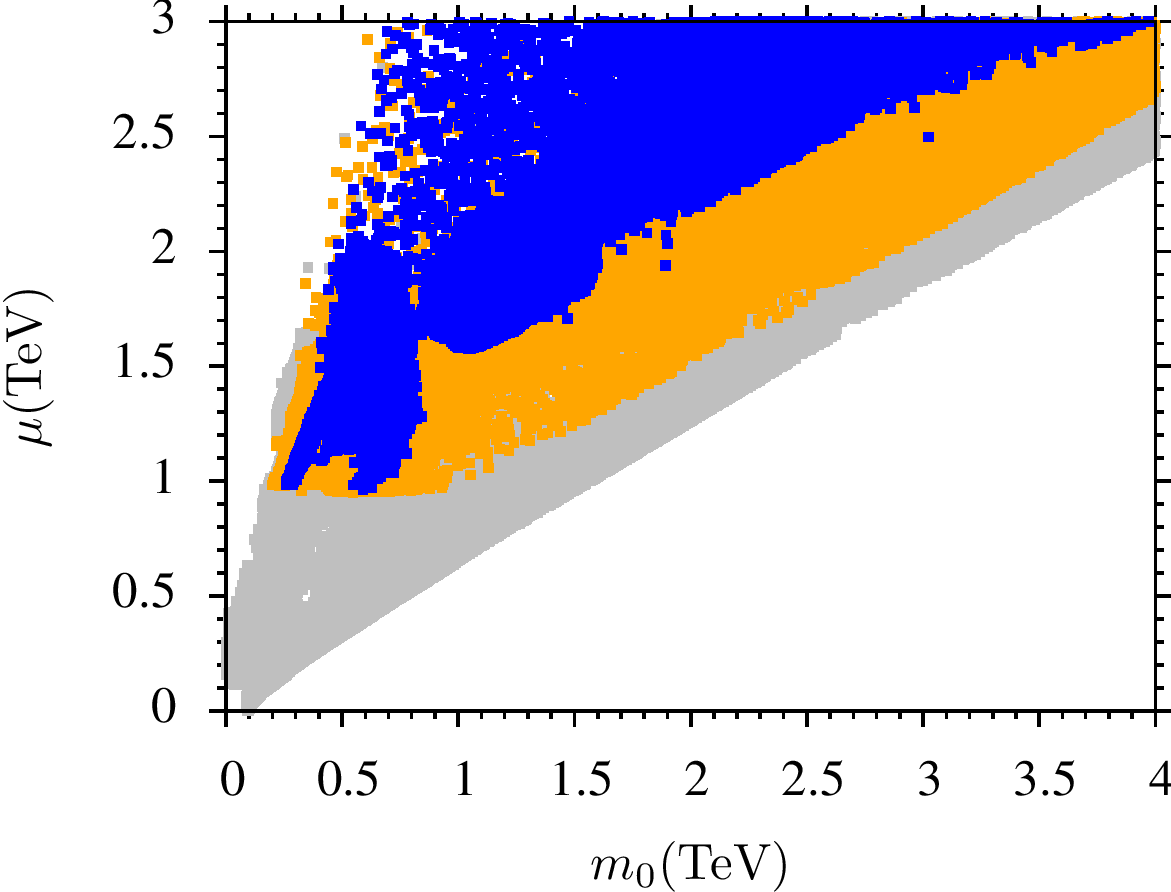}
}

\caption{
Plots in $m_{0}-\mu$ plane for CMSSM (left panel) and CMSSM-ISS (right panel).
Color coding is the same as in Figure~\ref{CMSSM_funda}.
}
\label{CMSSM_mu}
\end{figure}
\begin{figure}[ht!]
\centering
\subfiguretopcaptrue

\subfigure{
\includegraphics[totalheight=5.5cm,width=7.cm]{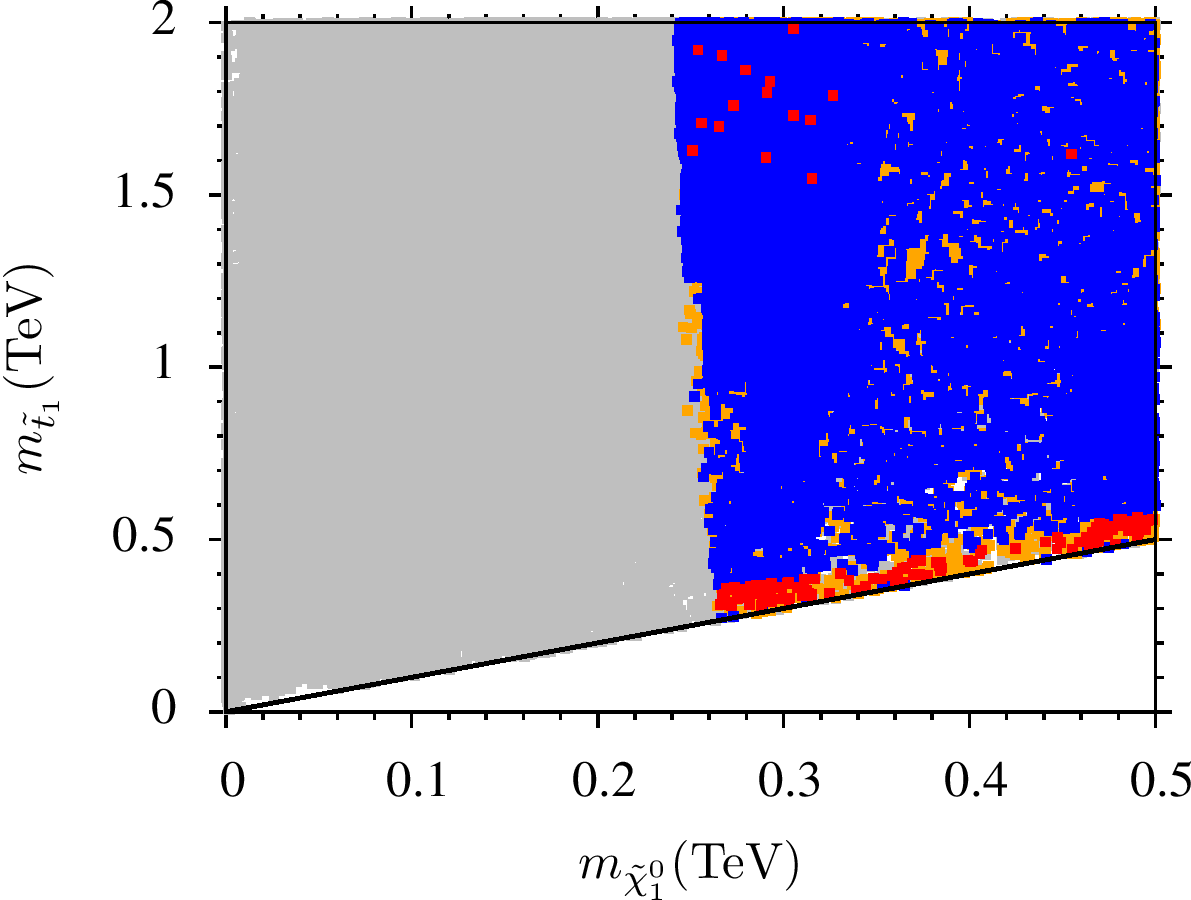}
}
\subfigure{
\includegraphics[totalheight=5.5cm,width=7.cm]{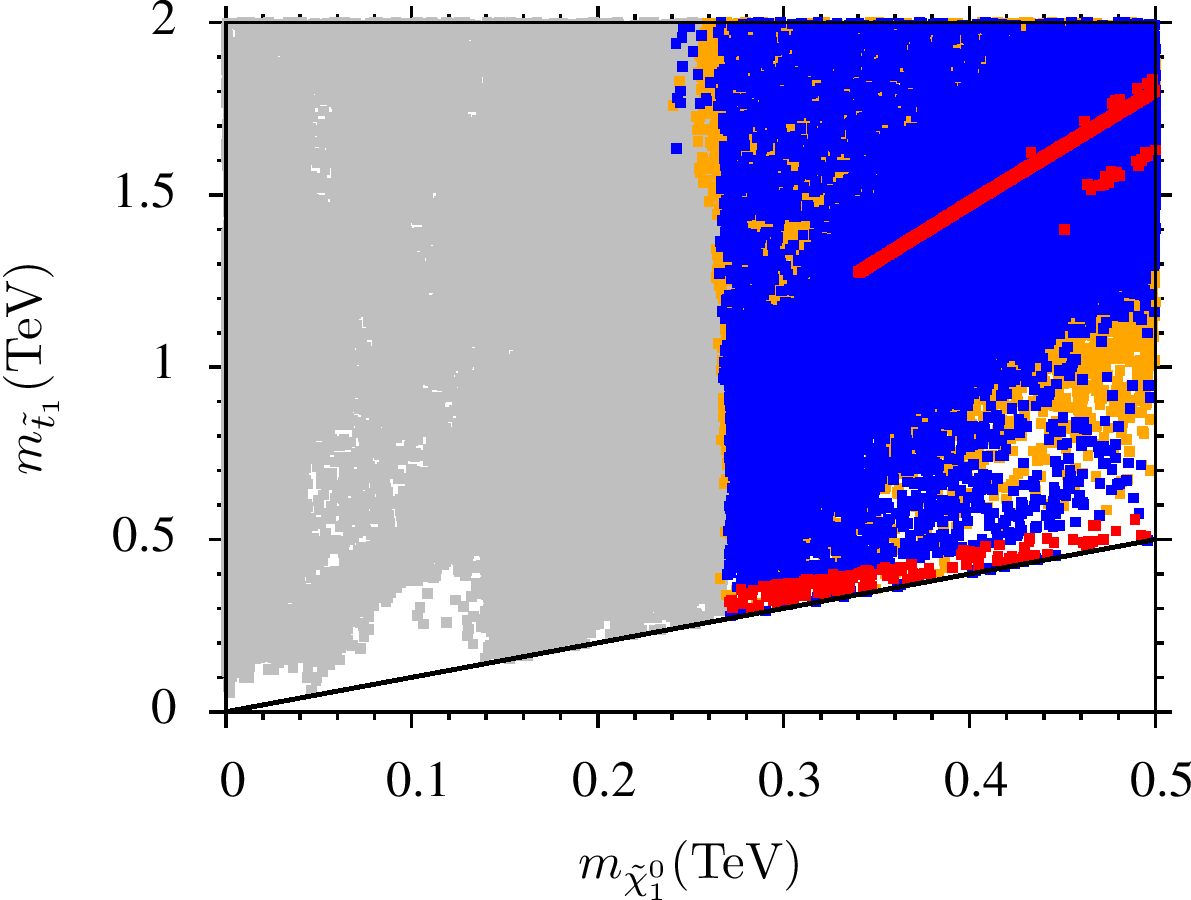}
}
\subfigure{
\includegraphics[totalheight=5.5cm,width=7.cm]{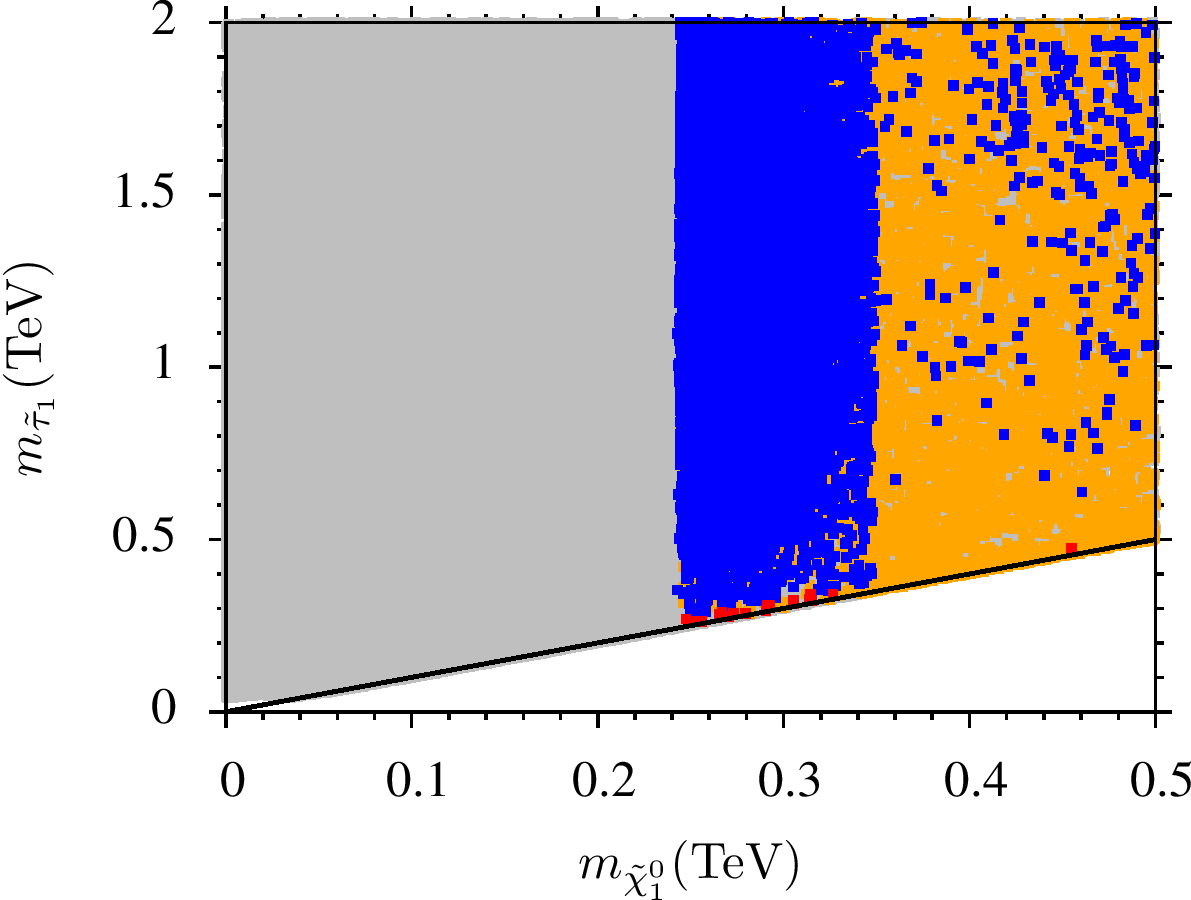}
}
\subfigure{
\includegraphics[totalheight=5.5cm,width=7.cm]{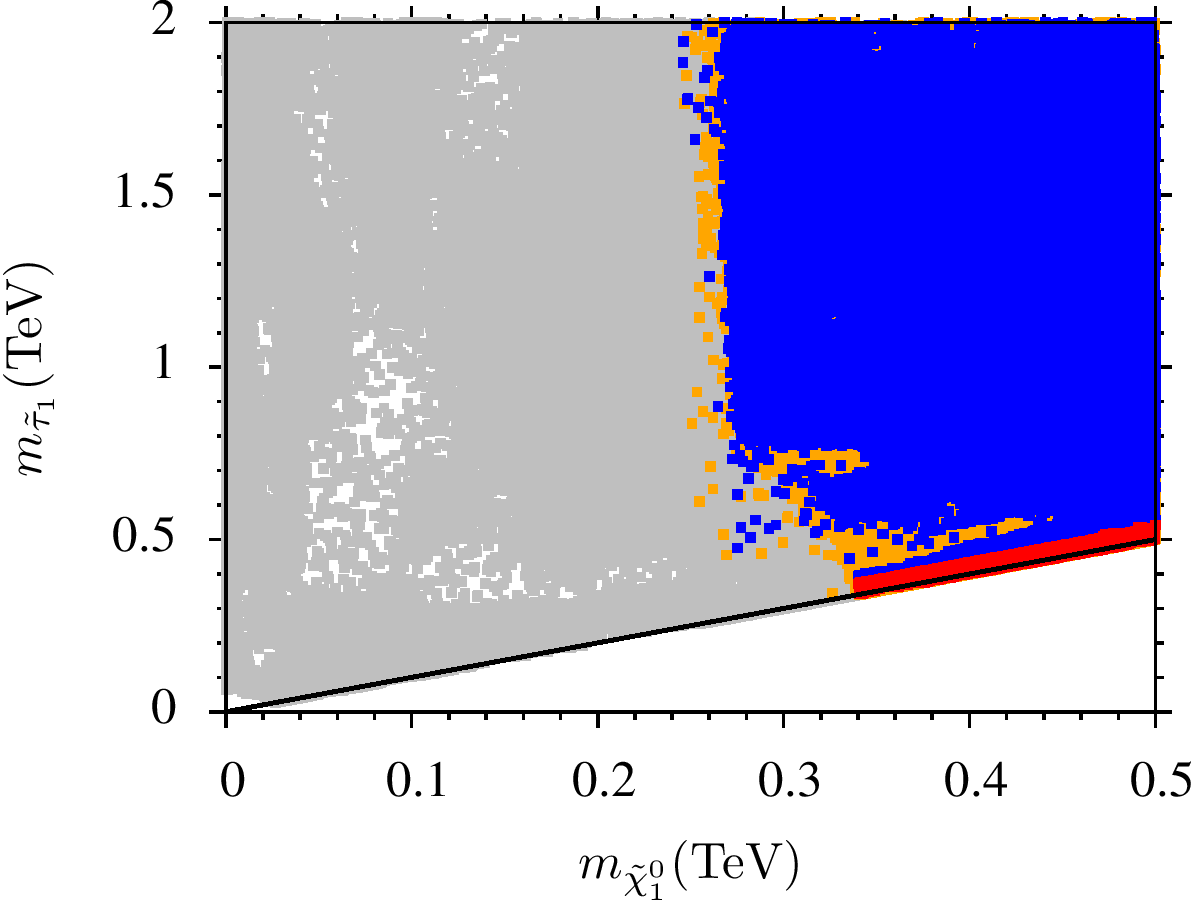}
}

\caption{
Plots in $m_{\tilde \chi_{1}^{0}}-m_{\tilde t_1}$ and $m_{\tilde \chi_{1}^{0}}-m_{\tilde \tau_1}$ planes for
CMSSM (left panel) and CMSSM-ISS (right panel).
The color coding is the same as in Figure~\ref{CMSSM_funda} except that red points are a subset of blue point
solutions and also satisfy bounds for relic  abundance, $0.001 \leq \Omega h^2 \leq 1$.
}
\label{CMSSM_spec1}
\end{figure}
\begin{figure}[ht!]
\centering
\subfiguretopcaptrue

\subfigure{
\includegraphics[totalheight=5.5cm,width=7.cm]{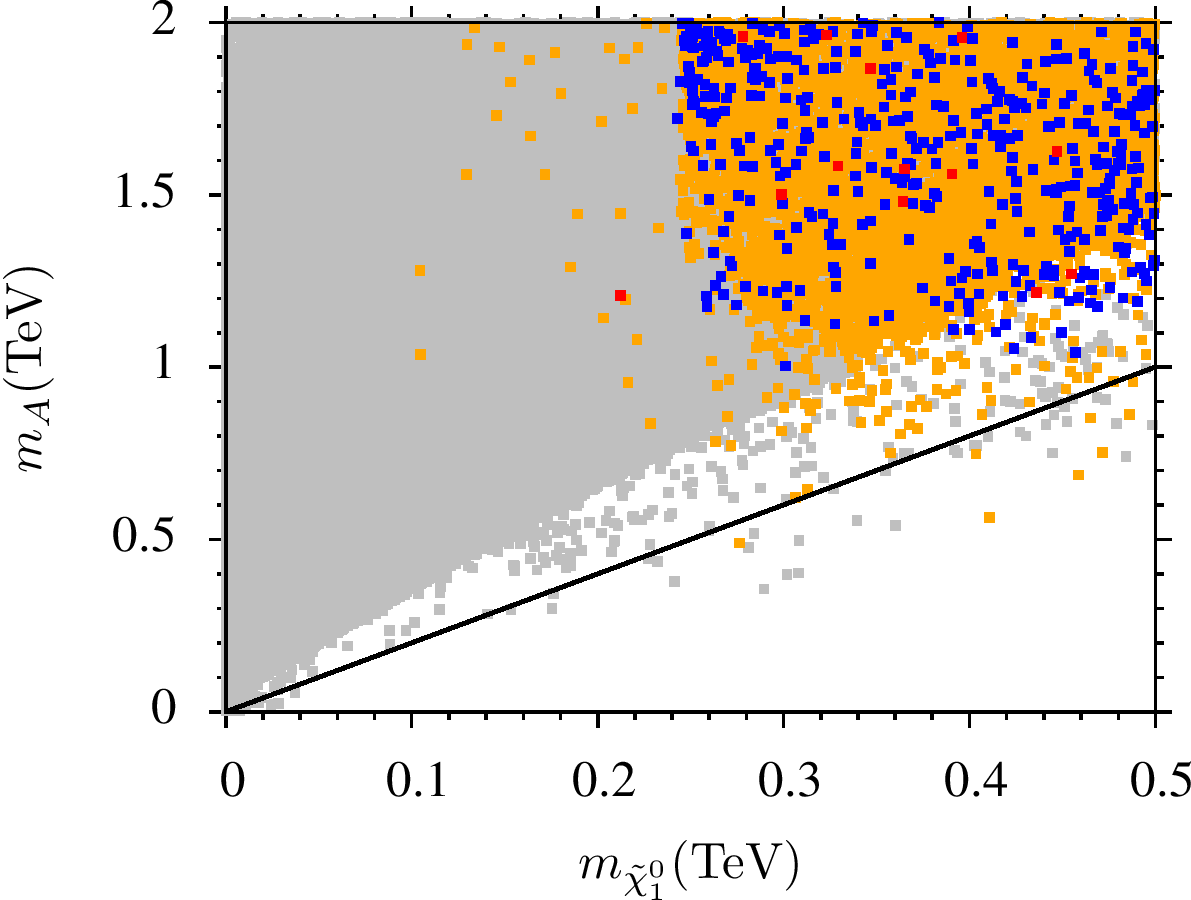}
}
\subfigure{
\includegraphics[totalheight=5.5cm,width=7.cm]{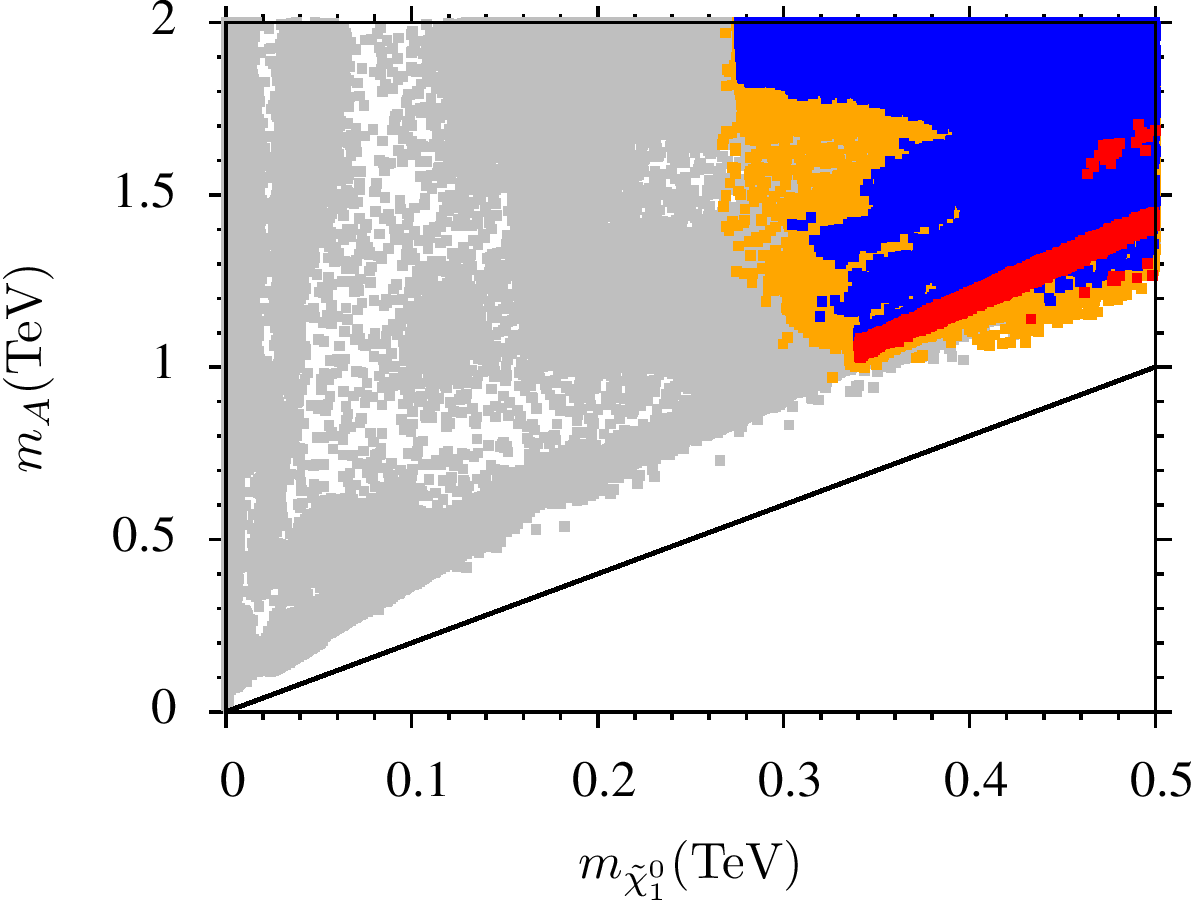}
}
\subfigure{
\includegraphics[totalheight=5.5cm,width=7.cm]{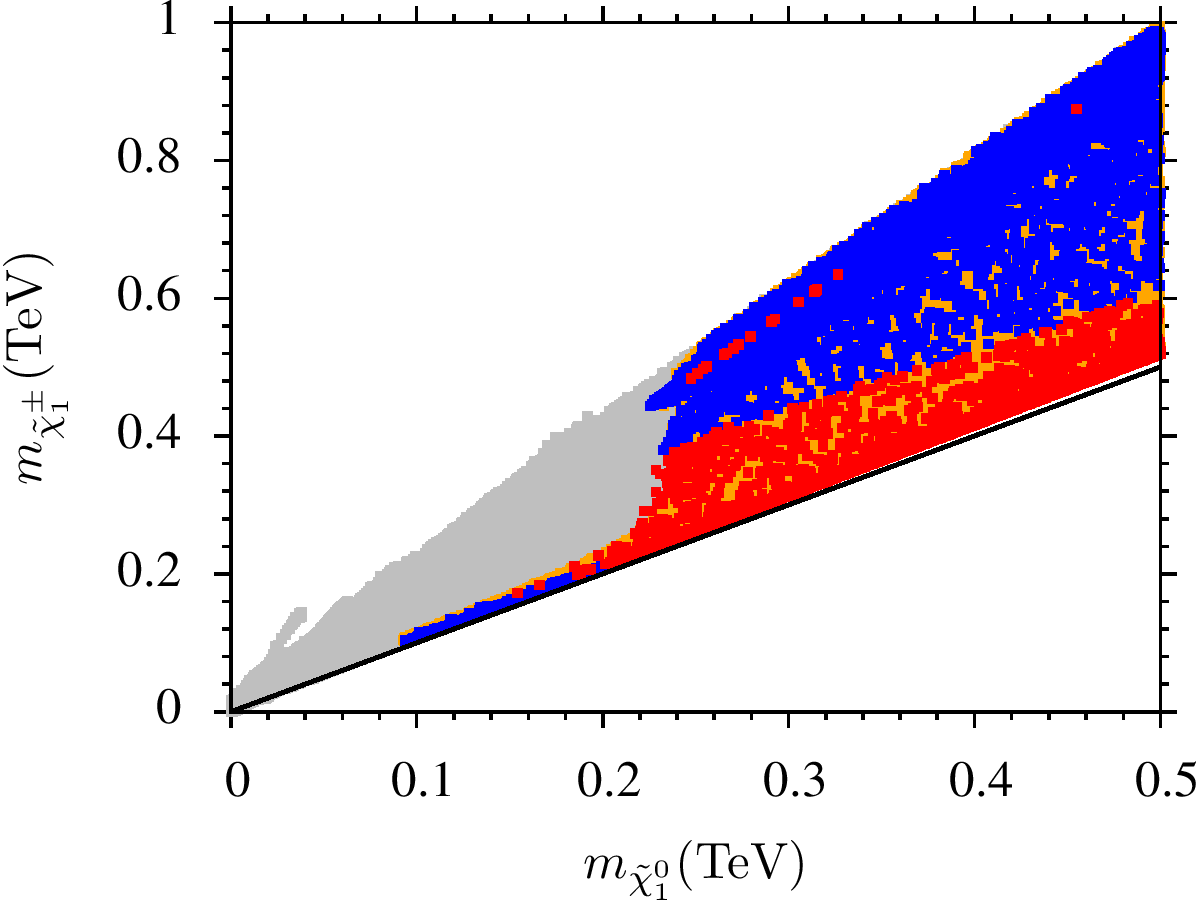}
}
\subfigure{
\includegraphics[totalheight=5.5cm,width=7.cm]{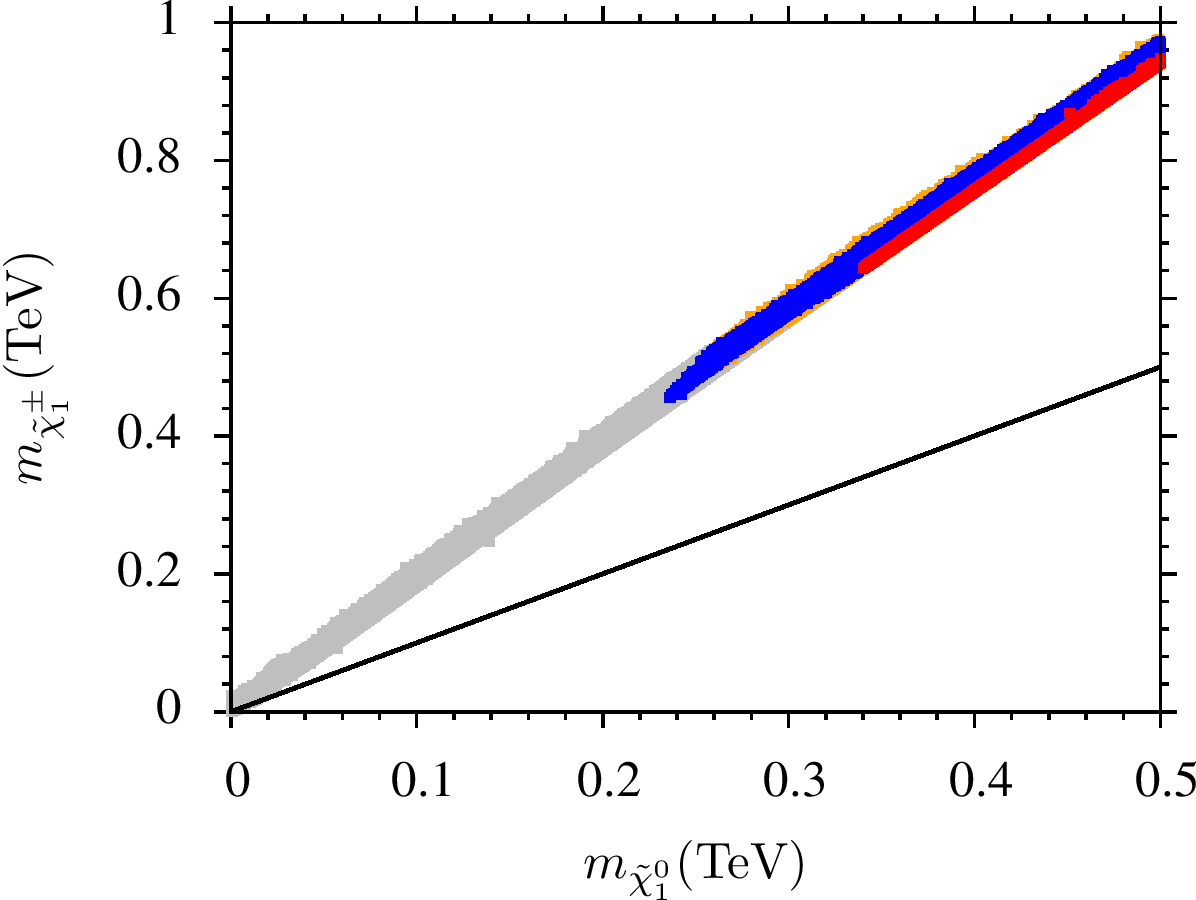}
}

\caption{
Plots in $m_{\tilde \chi_{1}^{0}}-m_{A}$ and $m_{\tilde \chi_{1}^{0}}-m_{\tilde \chi_{1}^{\pm}}$ planes for
CMSSM (left panel) and CMSSM-ISS (right panel).
The color coding is the same as in Figure~\ref{CMSSM_funda} except that red points are subset of blue point
solutions and also satisfy bounds for relic  abundance, $0.001 \leq \Omega h^2 \leq 1$.
}
\label{CMSSM_spec2}
\end{figure}


\begin{figure}[h]
\centering
\subfiguretopcaptrue

\subfigure{
\includegraphics[width=8cm]{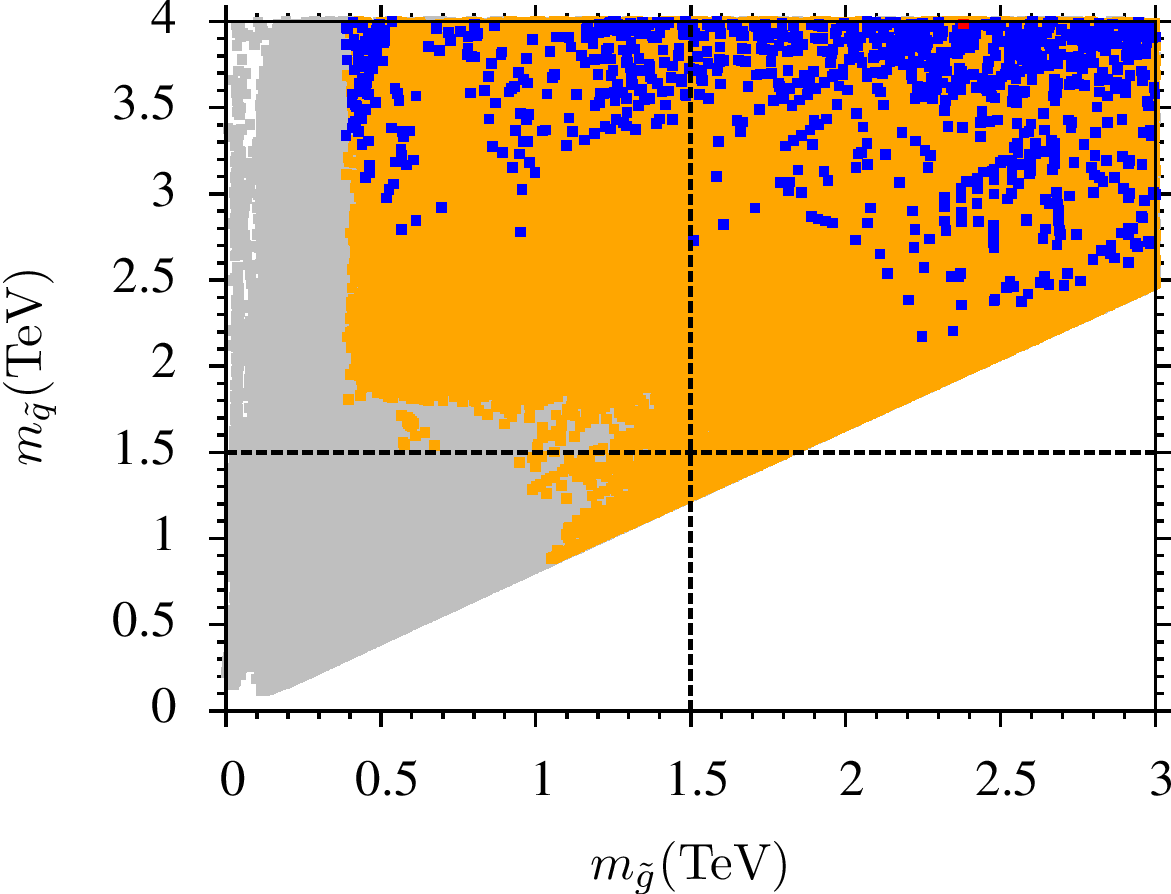}
}
\subfigure{
\includegraphics[width=8cm]{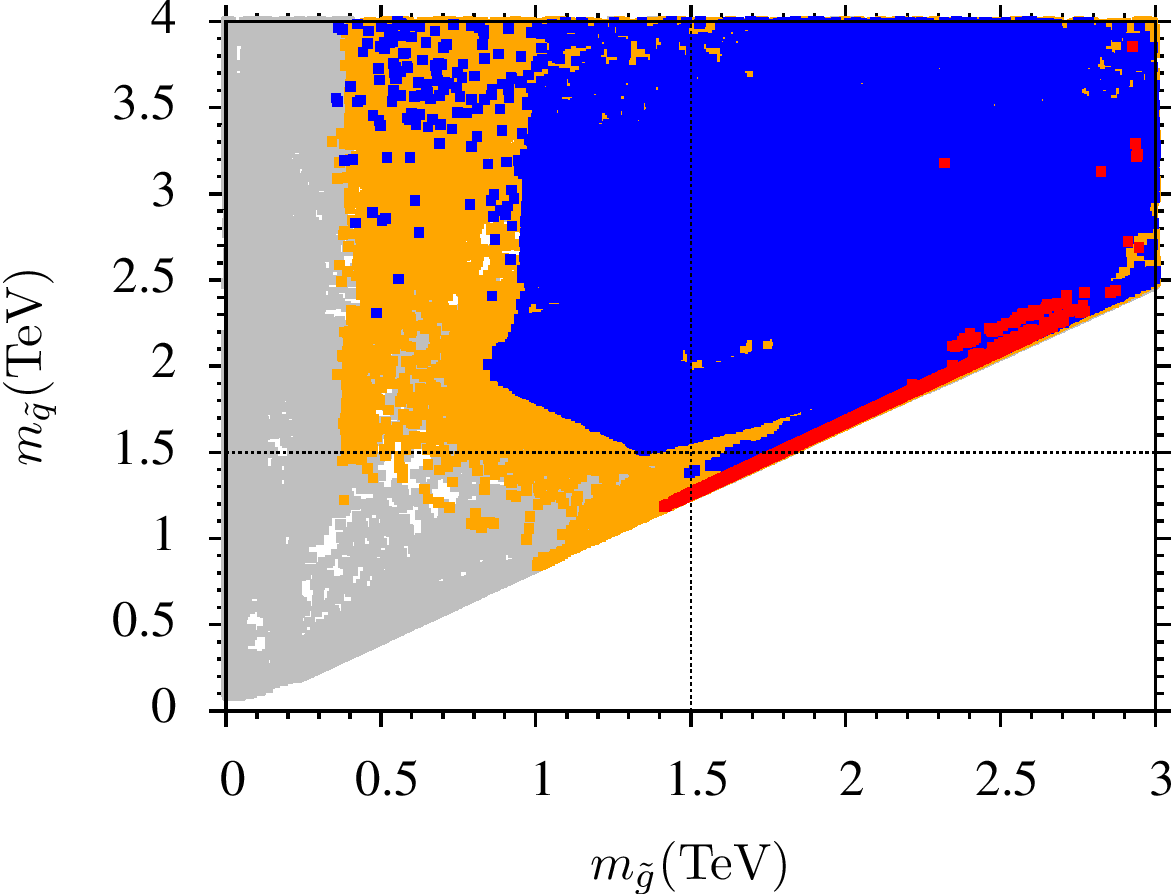}
}
\caption{
Plots in $m_{\tilde{g}}- m_{\tilde{q}}$ planes for CMSSM (left panel) and CMSSM-ISS (right panel).
The color coding is the same as in Figure~\ref{CMSSM_funda}, except that orange points do not
satisfy mass bounds for gluinos and first two generation squarks, and red points are a subset of
blue point solutions and also satisfy bounds for relic abundance, $0.001 \leq \Omega h^2 \leq 1$.
Dashed vertical and horizontal lines stand for current squark and gluino lower mass bounds respectively.
}
\label{fig56}
\end{figure}

\begin{figure}[h]
\centering
\subfiguretopcaptrue
\subfigure{
\includegraphics[width=8cm]{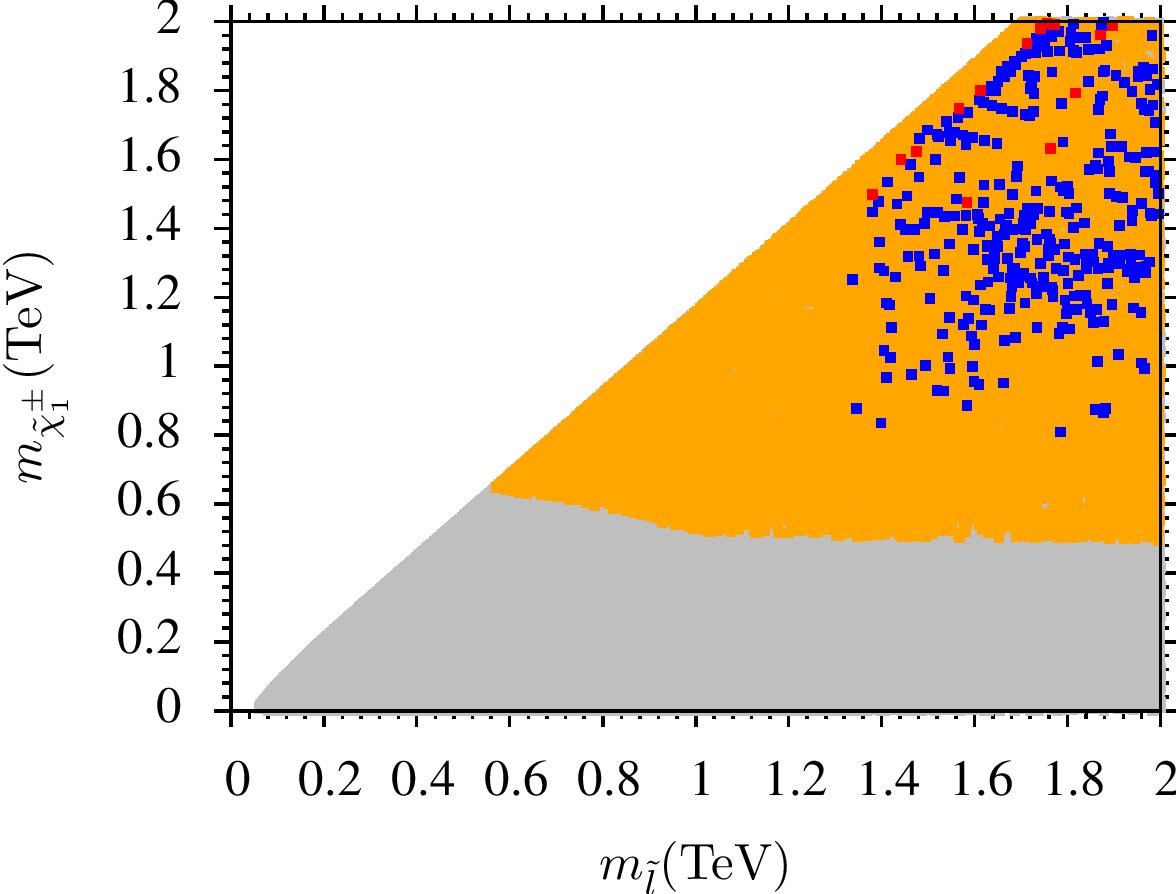}
}
\subfigure{
\includegraphics[width=8cm]{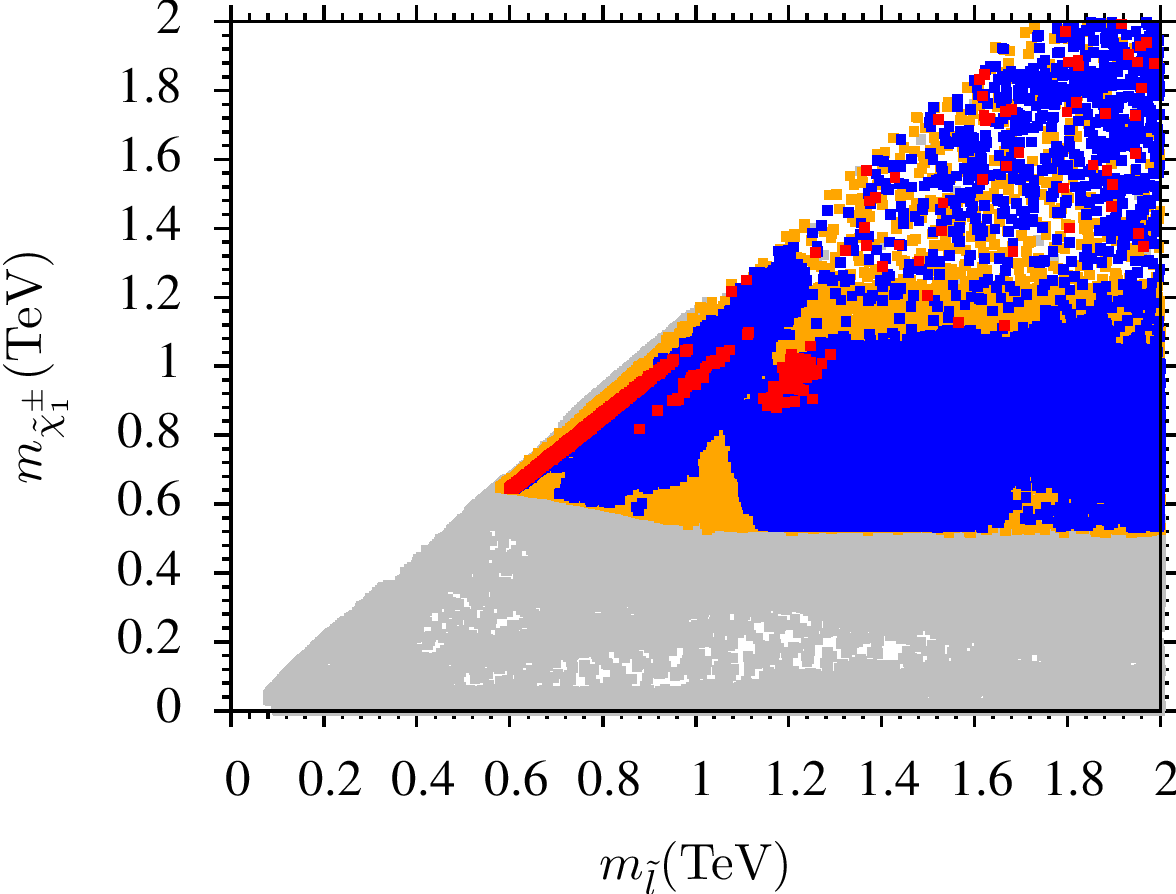}

}

\caption{
Plots in $m_{\tilde{l}}-m_{\tilde \chi_{1}^{\pm}}$ planes for CMSSM (left panel) and CMSSM-ISS (right panel).
The color coding is the same as in Figure~\ref{CMSSM_funda}, except that red points are a subset of blue point
solutions and also satisfy bounds for relic abundance, $0.001 \leq \Omega h^2 \leq 1$.
}
\label{fig55}

\end{figure}
\begin{table}[h!]
\centering
\begin{tabular}{lcc}
\hline
\hline
               & Point A  & Point B   \\
\hline
$m_{0}$        & 1020.3  & 3234     \\
$M_{1/2} $      & 1091.1   & 684.6      \\
$A_0/m_0$       & -2.71 & -2.97  \\
$\tan\beta$     & 38    & 14.4    \\
\hline
$m_h$          & 125  & 125  \\
$m_H$          & 1602 & 4744   \\
$m_A$          & 1592 & 4714   \\
$m_{H^{\pm}}$  & 1604 & 4745 \\
$\mu$          & 1772 & 3727 \\
\hline
$m_{\tilde{g}}$ & 2401  & 1705         \\
$m_{\tilde{\chi}^0_{1,2}}$
                & \textbf{476}, 902  & \textbf{312}, 608  \\
$m_{\tilde{\chi}^0_{3,4}}$
               & 1769, 1772   & 3724, 3724     \\

$m_{\tilde{\chi}^{\pm}_{1,2}}$
               & 905, 1773   & 614, 3733    \\
\hline $m_{ \tilde{u}_{L,R}}$
               & 2391, 2314   & 3492, 3476    \\
$m_{\tilde{t}_{1,2}}$
               & 1569, 1983  & \textbf{347}, 2376  \\
\hline $m_{ \tilde{d}_{L,R}}$
               & 2392, 2305   & 3493, 3479      \\
$m_{\tilde{b}_{1,2}}$
               & 1940, 2035   & 2400, 3262    \\
\hline
$m_{\tilde{\nu}_{1}}$
              & 1248   & 3265       \\
$m_{\tilde{\nu}_{3}}$ & 792   & 2024         \\
\hline
$m_{ \tilde{e}_{L,R}}$
              & 1252, 1098  & 3261, 3245   \\
$m_{\tilde{\tau}_{1,2}}$
             & \textbf{497}, 820   & 2040, 3027   \\
\hline
\hline
$\sigma_{SI}({\rm pb})$
             & 1.57$\times 10^{-11}$  & 1.71$\times 10^{-15}$  \\
$\sigma_{SD}({\rm pb})$
             & 5.05$\times 10^{-9}$  & 7.3$\times 10^{-13}$  \\
$\Omega_{CDM}h^{2}$
            & 0.114  & 0.092  \\
\hline
\hline
\end{tabular}
\caption{
Masses (in GeV units) and other parameters for two CMSSM-ISS
 benchmark points satisfying all phenomenological constraints discussed in section \ref{sec:scan}.
Points A and B are chosen from
 the stau-neutralino coannihilation and the stop-neutralino coannihilation regions respectively.
\label{tab:CMSSM_iss}}
\end{table}

\section{Results}
\label{results}
\subsection{CMSSM and Inverse Seesaw}
\label{CMSSM}

In this section we present our results for the CMSSM and the CMSSM with additional ISS contribution (CMSSM-ISS).
The main idea behind the presentation of these results is to show that these two scenarios have quite distinct
features as far as choice for the fundamental parameters
of the models is concerned. In Figure~\ref{CMSSM_funda}, the left panels represent our results for the CMSSM, while the right panels display
our results for the CMSSM-ISS. Here grey points satisfy REWSB and the LSP neutralino requirement.
The orange points represent solutions which satisfy the mass bounds and B-physics bounds from Section~\ref{sec:scan}.
Solutions in blue color are a subset of orange points and satisfy the requirement
 $123\, {\rm GeV}\lesssim m_h\lesssim 127\, {\rm GeV}$.
 This figure clearly serves our purpose stated above.

For instance, the graph in $m_{0}-m_{1/2}$ plane shows that for the CMSSM  case, the Higgs mass bounds excludes
simultaneously small
values for  $m_{0}$ and $m_{1/2}$, while in the CMSSM-ISS case, we can have relatively small values for  $m_{1/2}$ ($<$ 800~GeV) and $m_0$ ($<$ 400~GeV), consistent with all
constraints given in section~\ref{sec:scan}.
There is also noticeable difference between CMSSM and CMSSM-ISS in the $A_{0}/m_{0}-m_{0}$ plane.
In the CMSSM
case, for instance, we find $m_0 \sim 700$~GeV for $A_0/m_0 =-3$, and for $A_0/m_0=3$ we have
$m_0 \sim 1.3$~TeV. In CMSSM-ISS, on the other hand, the corresponding minimum $m_0$ values vary from 400~GeV
to 1.1~TeV.
In the $m_{0}-\tan\beta$ plane too, considering the blue points, we see in the left panel that for a minimum value
$m_0 \sim 700$~GeV, the corresponding $\tan\beta$ value is around 16. In the right panel, on the other hand, $\tan\beta$ is again around 16 but now the minimum value of $m_0$ is $\sim 300$~GeV.

In Figure~\ref{CMSSM_mu} we show plots of  $m_{0}$ versus $\mu$. The color coding is the same as in
Figure~\ref{CMSSM_funda} with the left and right panels representing CMSSM  and CMSSM-ISS respectively. This figure shows
very distinct features of the two scenarios. Considering the orange points, in CMSSM-ISS we have solutions with $\mu\gtrsim 1\,{\rm~TeV}$, in
contrast with the CMSSM, where we have solutions with small, as well as large values of $\mu$.
The reason for this difference is that in CMSSM-ISS, $m_{H_u}^2$ gets new contribution from the loop induced
by the coupling $N^c_i H_u L_j$ in addition to the top quark loop, which makes $\mu$ relatively heavy.
Thus, in the CMSSM-ISS case we do not have the so-called focus point/hyperbolic branch
scenario~\cite{focus, Chan:1997bi} while it is still a viable solution in the CMSSM case.

In Figure~\ref{CMSSM_spec1}, we show graphs in $m_{\tilde \chi_{1}^{0}}-m_{\tilde t_1}$ and $m_{\tilde \chi_{1}^{0}}-m_{\tilde \tau_1}$ planes. The color coding is the same as in Figure~\ref{CMSSM_funda}, except that the solutions in red are a subset of solutions in blue and also satisfy the relic  abundance bound $0.001 \leq \Omega h^2 \leq 1$.
These graphs show that despite the fact that there are differences in the
space of fundamental parameters, the mass spectrum for $\chi_{1}^{0}$, $\tilde t_1$ and $\tilde \tau_1$ turn out to be more or less identical.

For instance, in the $m_{\tilde \chi_{1}^{0}}-m_{\tilde t_1}$
plane we see that we have NLSP $\tilde t_1$ in the mass range of $\sim 260-500 \,{\rm~GeV}$ in both cases.
Similar results were also reported in
\cite{Baer:2012by,Gogoladze:2011be} in the case of $b$-$\tau$ Yukawa coupling unification in CMSSM and $SU(5)$.
It was shown in \cite{Ajaib:2011hs,He} that the region of parameter space with stop-neutralino mass
difference of 20$\%$ is ruled out for
$m_{\tilde t_1} \lesssim 140\,{\rm~GeV}$. In the $m_{\tilde \chi_{1}^{0}}-m_{\tilde \tau_1}$ plane, we note that NLSP $\tilde \tau_1$
 has the same mass range in CMSSM and
CMSSM-ISS. The reason why we have comparable intervals for $m_{\tilde t_1}$ and $m_{\tilde \tau_1}$ in  CMSSM and CMSSM-ISS
is that low values for both sparticle masses are acheived via
fine tuning involving the trilinear SSB terms, while
the addition of ISS to CMSSM mostly affects the first two generation sparticle masses.

In Figure~\ref{CMSSM_spec2}, we present graphs in $m_{\tilde \chi_{1}^{0}}-m_{A}$ and
$m_{\tilde \chi_{1}^{0}}-m_{\tilde \chi_{1}^{\pm}}$ planes, with color coding the same as in Figure~\ref{CMSSM_spec1}. The graphs in
$m_{\tilde \chi_{1}^{0}}-m_{A}$ plane show that we do not have the
$A$-resonance solution~\cite{funnel}, and the reason can be understood from the following equation:
\beq
m_{A}^2=2|\mu|^2 +m_{H_u}^{2}+m_{H_d}^2.
\label{mA}
\eeq
In CMSSM, since we have universal scalar masses and we require  $m_h \sim 123-127 \,{\rm~GeV}$, $m_{H_u}^2$ and
$m_{H_d}^2$ are both large, and, as a result, $m_A$ is also large. This can be seen in the
$m_{\tilde \chi_{1}^{0}}-m_{A}$ graph in the left panel.
The solid black line in the graph represents the condition $2m_{\tilde \chi_{1}^{0}}=m_{A}$ for the $A$-resonance
solution~\cite{funnel}.

We note that the solutions in orange  color lie around the solid black line, but if we apply the constraint $123 \,{\rm~GeV} \lesssim m_h \lesssim 127 \,{\rm~GeV}$,
the relevant blue points lie further from the black line. In the right panel, which represents the CMSSM-ISS case,
we note that both the orange and blue points are further away from the solid black line. This is because of two reasons.
Firstly, as stated earlier, $\mu$ is larger because of extra contributions from the $N^c_i H_u L_j$
Yukawa coupling, and so the orange points move away from the solid black line.
Secondly, as explained, in the CMSSM case the $m_h$ constraint makes solutions move
away from the solid black line as $m_A$ becomes larger.

A more distinctive figure concerning the sparticle spectra in CMSSM and
CMSSM-ISS is presented in
the $m_{\tilde \chi_{1}^{0}}-m_{\tilde \chi_{1}^{\pm}}$ plane. In contrast to CMSSM (left panel), the figure for CMSSM-ISS is quite different. This is due to the fact that in CMSSM-ISS, the LSP neutralino is mostly a bino and the chargino mostly wino.
Therefore, the ratio $m_{\tilde \chi_{1}^{0}} / m_{\tilde \chi_{1}^{\pm}}$ is close to
the ratio of $U(1)$ and $SU(2)$ gauge couplings, $g_1/g_2\thickapprox1/2$, and the points form a narrow strip.

In Figure~\ref{fig56} we show  $m_{\tilde{q}}$ versus $m_{\tilde{g}}$  for CMSSM (left panel) and CMSSM-ISS
(right panel). The color coding is the same as in Figure~\ref{CMSSM_funda}, except that the orange points
do not include mass bounds for gluinos and the first two generation squarks.
Dashed vertical and horizontal lines represent current squark and gluino mass bounds.
We note that especially in the CMSSM the gluino mass bound excludes a significant portion of the parameter space
which otherwise is consistent with the experimental data. The location of blue points relative to the orange points
shows how the lower bounds on the squark and gluino masses are pushed up by $m_h$. It is interesting to observe that there are no red points with  neutralino LSP dark matter within the reach of LHC14. Comparing results from
$m_{\tilde{g}}- m_{\tilde{q}}$ panel with the results from Figures \ref{CMSSM_spec1}
and \ref{CMSSM_spec2},  we conclude that in the CMSSM, the solution which yields the correct dark matter
relic abundance predicts gluino and squarks masses that lie beyond the reach of the LHC14~\cite{cms_lim}.

On the other hand, comparison of left and right panels in Figure~\ref{fig56} shows the impact of the
ISS mechanism on the sparticle masses. We can see from the
$m_{\tilde{g}}- m_{\tilde{q}}$ plot in the right panel that plenty of blue points are left after we apply the Higgs mass constraint  $123 \,{\rm~GeV} \lesssim m_h \lesssim 127 \,{\rm~GeV}$.
This means that in the presence of the ISS mechanism, most points satisfying all experimental constraints lie in the Higgs mass range $123 \,{\rm~GeV} \lesssim m_h \lesssim 127 \,{\rm~GeV}$, which is very different from the
CMSSM case. There are also red points in the right panel which shows that we can have LHC testable solutions with the correct
relic abundance of dark matter.

In Figure~\ref{fig55} we display plots for $m_{\tilde \chi_{1}^{\pm}}$ versus $m_{\tilde{l}}$ in CMSSM (left panel) and
CMSSM-ISS (right panel), with the color coding the same as in the previous figures.
In the left panel we see from the blue points that $m_{\tilde{l}}>$ 1.4~TeV,
which may be difficult to test at the LHC.  On the other hand,
we see in the right panel solutions in blue and red colors around $m_{\tilde{l}}\simeq 500$~GeV, which provides a glimmer of hope that sleptons
employing the CMSSM-ISS mechanism may be found at the LHC.

In Table~\ref{tab:CMSSM_iss} we display two benchmark points for the cMSM-ISS model that are consistent with constraints in
Section~\ref{sec:scan}.
The LSP neutralino relic density in the two cases is in accord with the WMAP observations, and corresponds to stau-neutralino~\cite{stauco} (stop-neutralino~\cite{dm:stop}) coannihilation for point A (B).
For point A, $m_{\tilde \tau_1} \approx 500$~GeV, $m_{\tilde g} \approx$ 2.4~TeV,
the first two generation squarks are close to 2~TeV, while slepton masses are around $1-2$ TeV.
For point B, $m_{\tilde t_1} \approx $ 350~GeV,  $m_{\tilde{g}} \approx 1.7$~TeV, the first two generation squark masses are about 3.4~TeV,
while slepton masses are around 3.2~TeV.

\subsection{NUHM2 and Inverse Seesaw}
\label{nuhm2}

\begin{figure}[ht!]
\centering
\subfiguretopcaptrue

\subfigure{
\includegraphics[totalheight=5.5cm,width=7.cm]{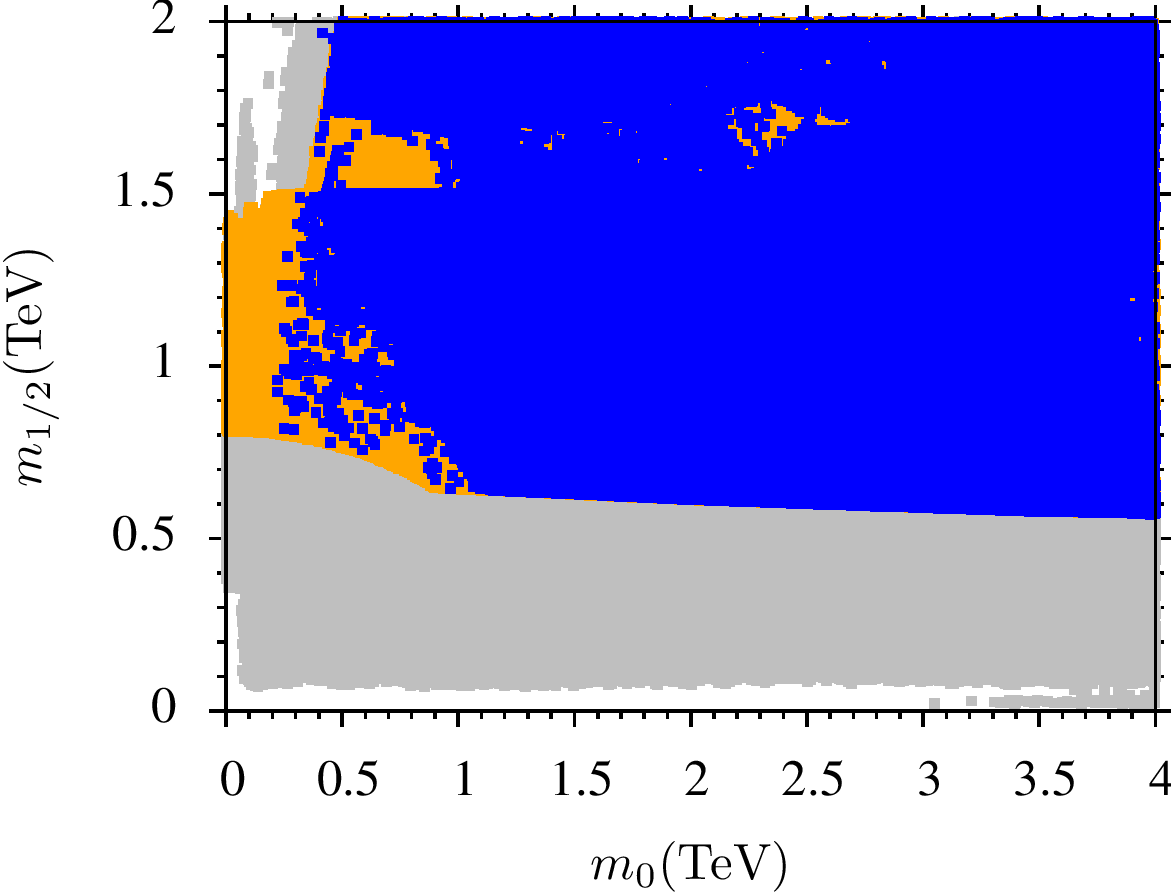}
}
\subfigure{
\includegraphics[totalheight=5.5cm,width=7.cm]{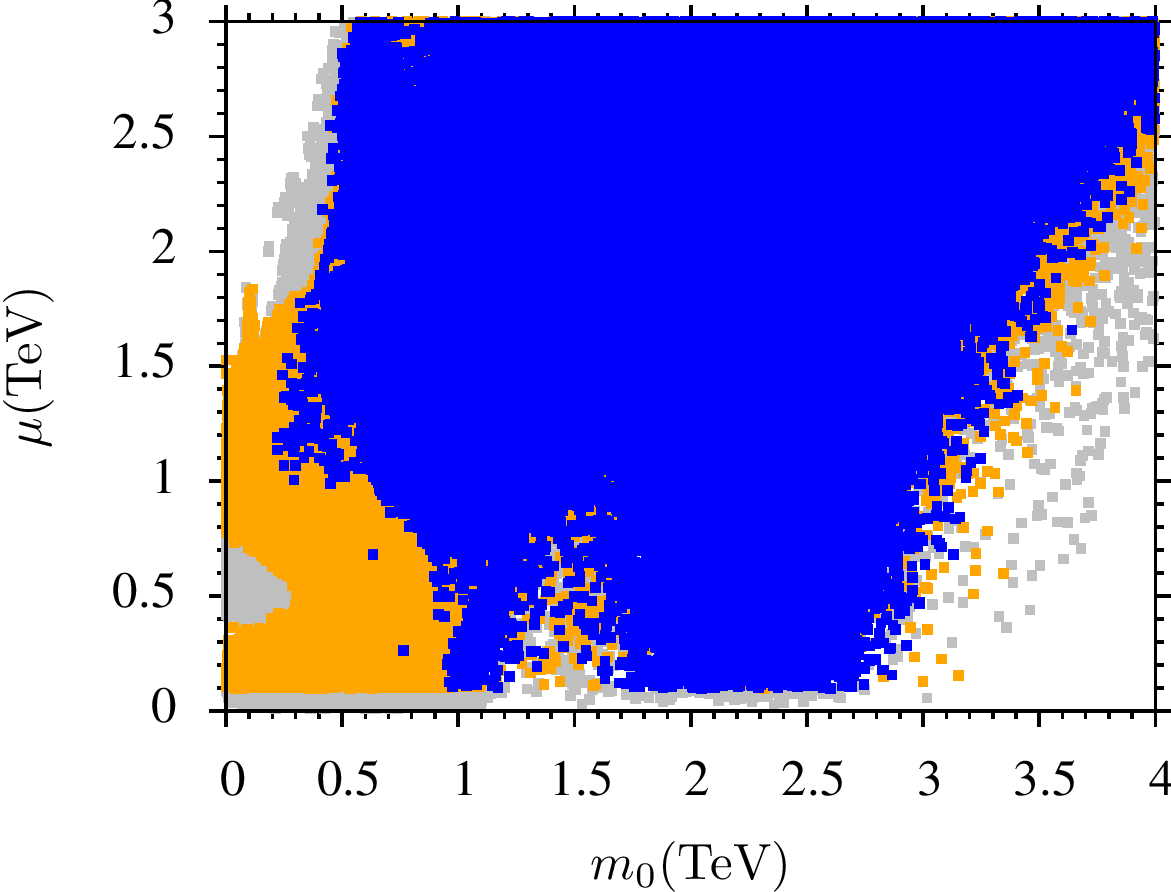}
}

\caption{
Plots in $m_{0}-m_{1/2}$  and $m_{0}-\mu$ planes for NUHM2-ISS.
The color coding is the same as in Figure~\ref{CMSSM_funda}.
}
\label{nuhm2_funda}
\end{figure}

In this subsection we present the results of our scan for NUHM2 with ISS contributions (NUHM2-ISS).
In Figure~\ref{nuhm2_funda}
we present graphs in $m_{0}-m_{1/2}$ and $m_{0}-\mu$ planes, with color coding the same as in Figure~\ref{CMSSM_funda}.
In the $m_{0}-m_{1/2}$ plane we see that the results are similar to what we found in CMSSM-ISS.
Again we can have solutions compatible with all experimental constraints presented in section~\ref{sec:scan}.
We note that the Higgs mass constraint $123\, {\rm GeV}\lesssim m_h \lesssim 127\, \rm{GeV}$ provides the lower bounds
$m_{1/2}\approx 500$~GeV and  $m_0\approx 1$~TeV. Since $\mu$ is a free parameter in NUHM2,
 we can find solutions with any value of $\mu$ compatible with the experimental
data (see  $m_{0}-\mu$ plot).
As shown in \cite{Gogoladze:2012yf}, a relatively small $\mu$ term is necessary, but not sufficient,
to be consistent with natural supersymmetry (little hierarchy problem) criteria.
We find that it is hard to fully resolve  the little hierarchy problem in this scenario.

\begin{figure}[ht!]
\centering
\subfiguretopcaptrue

\subfigure{
\includegraphics[totalheight=5.5cm,width=7.cm]{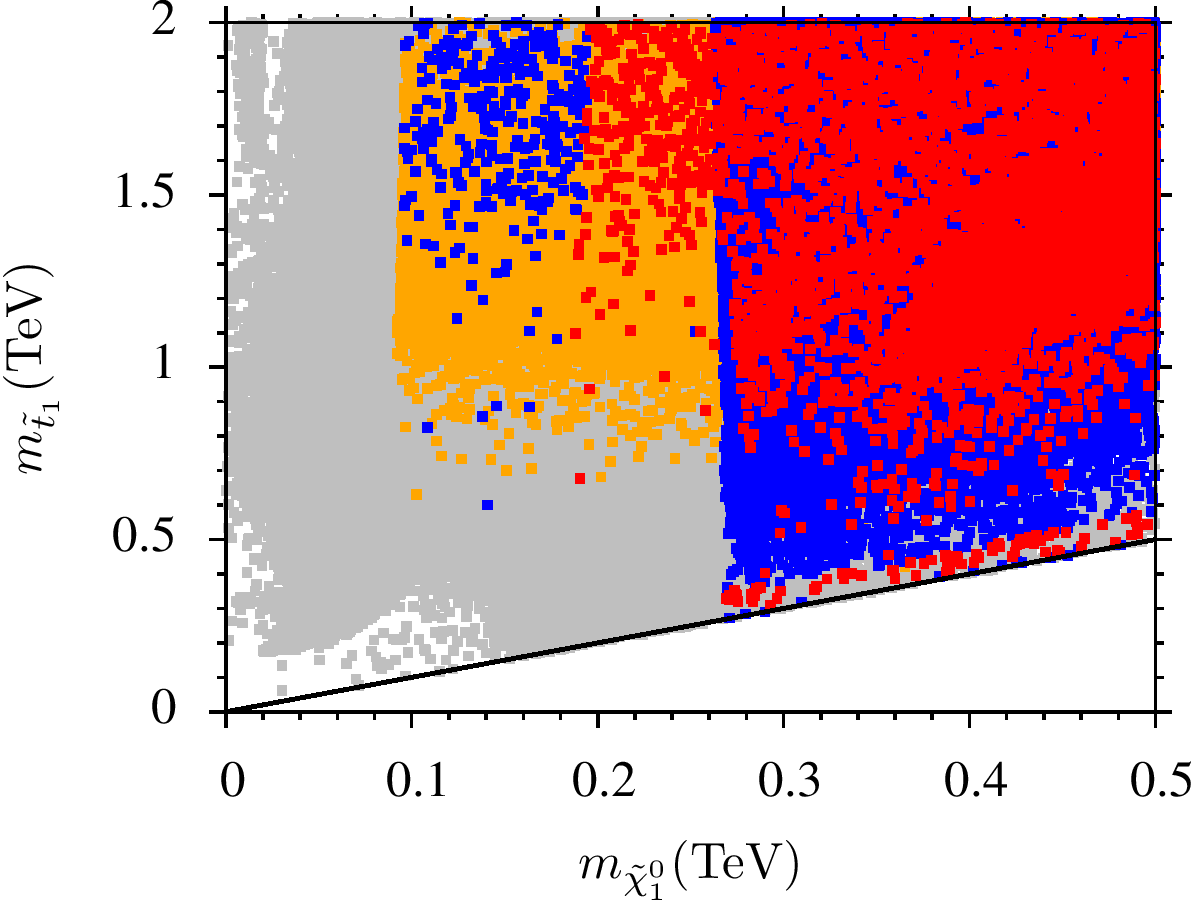}
}
\subfigure{
\includegraphics[totalheight=5.5cm,width=7.cm]{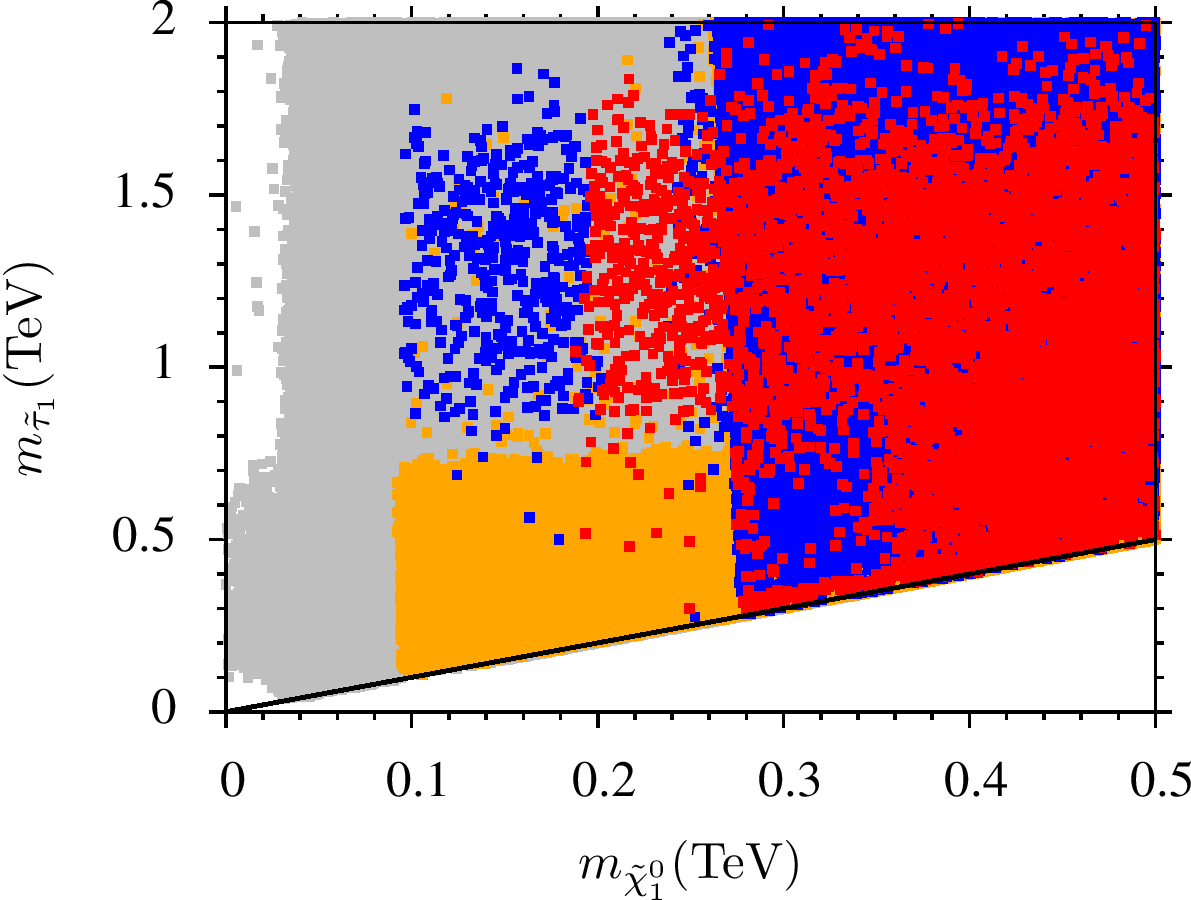}
}
\subfigure{
\includegraphics[totalheight=5.5cm,width=7.cm]{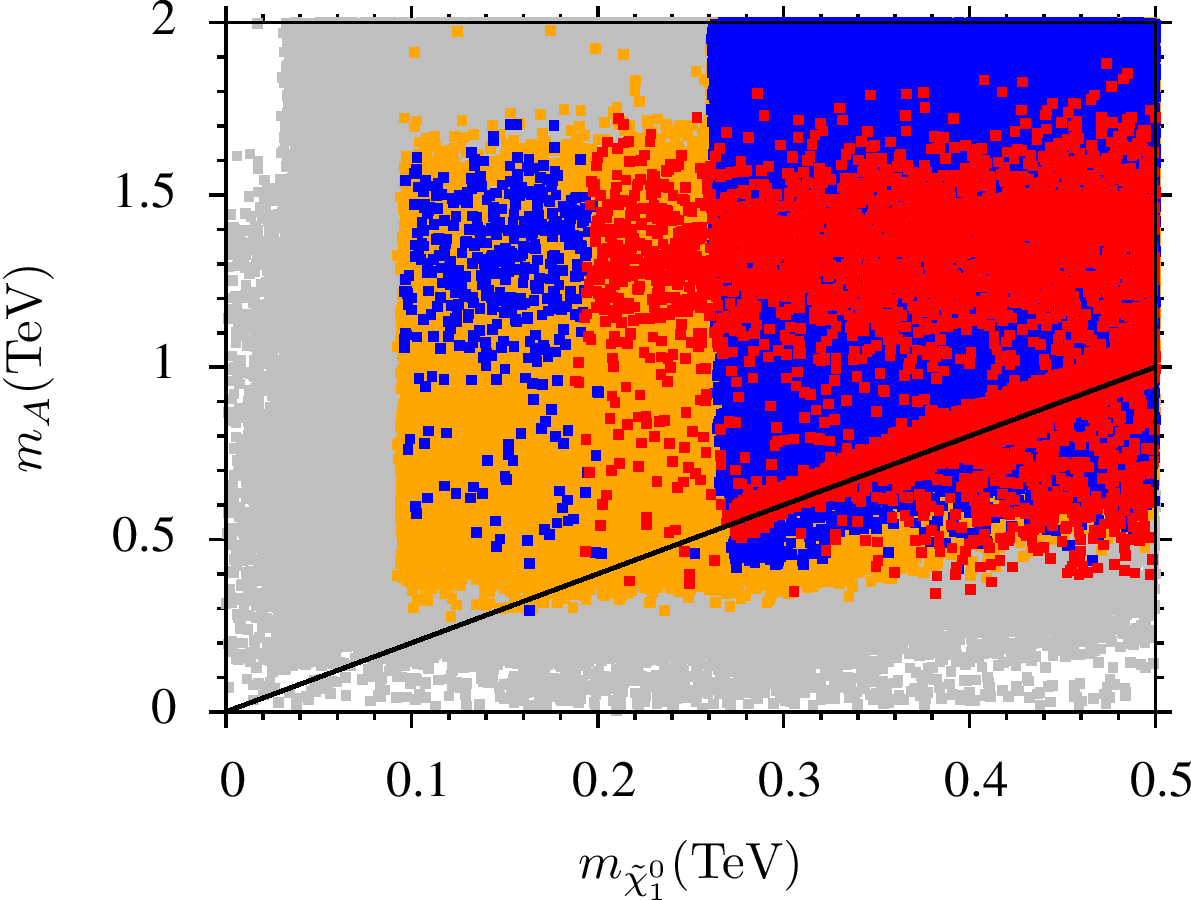}
}
\subfigure{
\includegraphics[totalheight=5.5cm,width=7.cm]{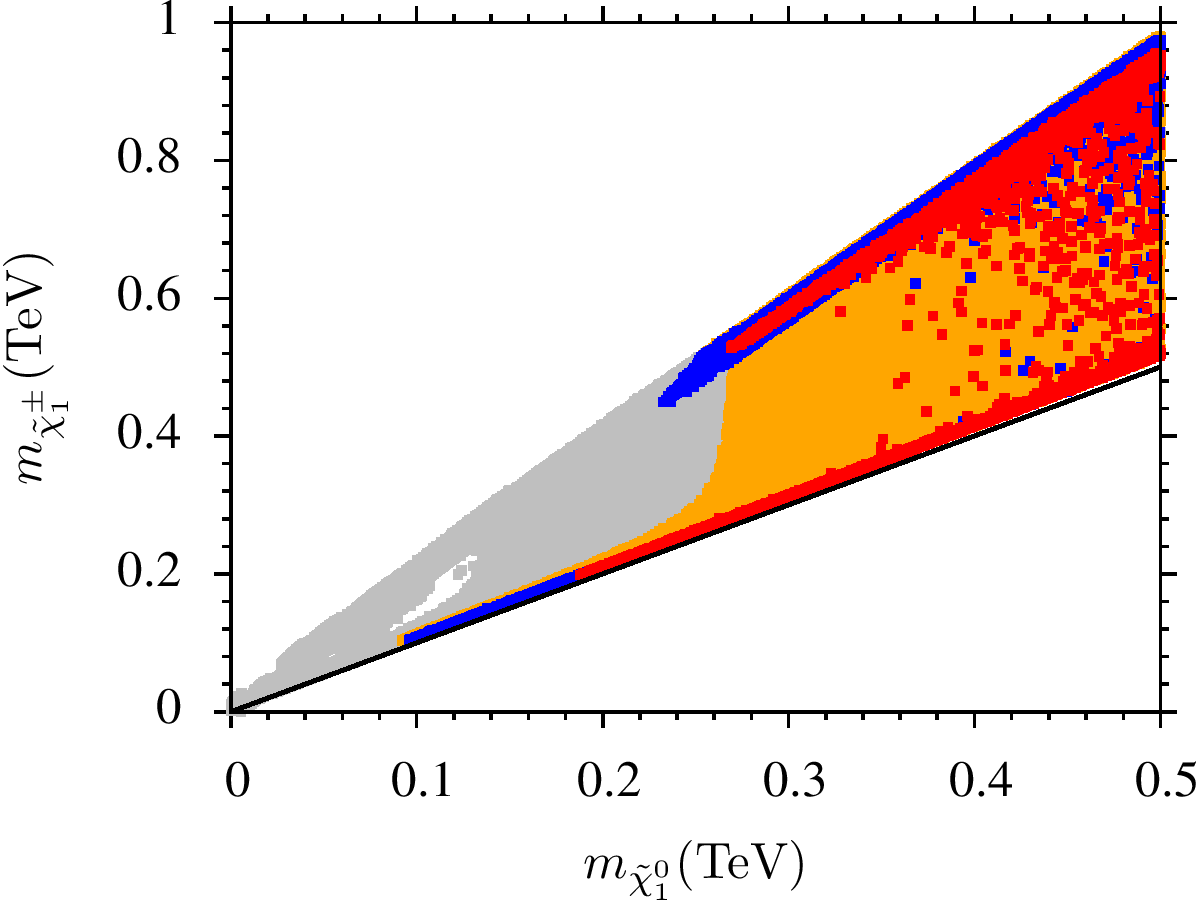}
}
\subfigure{
\includegraphics[totalheight=5.5cm,width=7.cm]{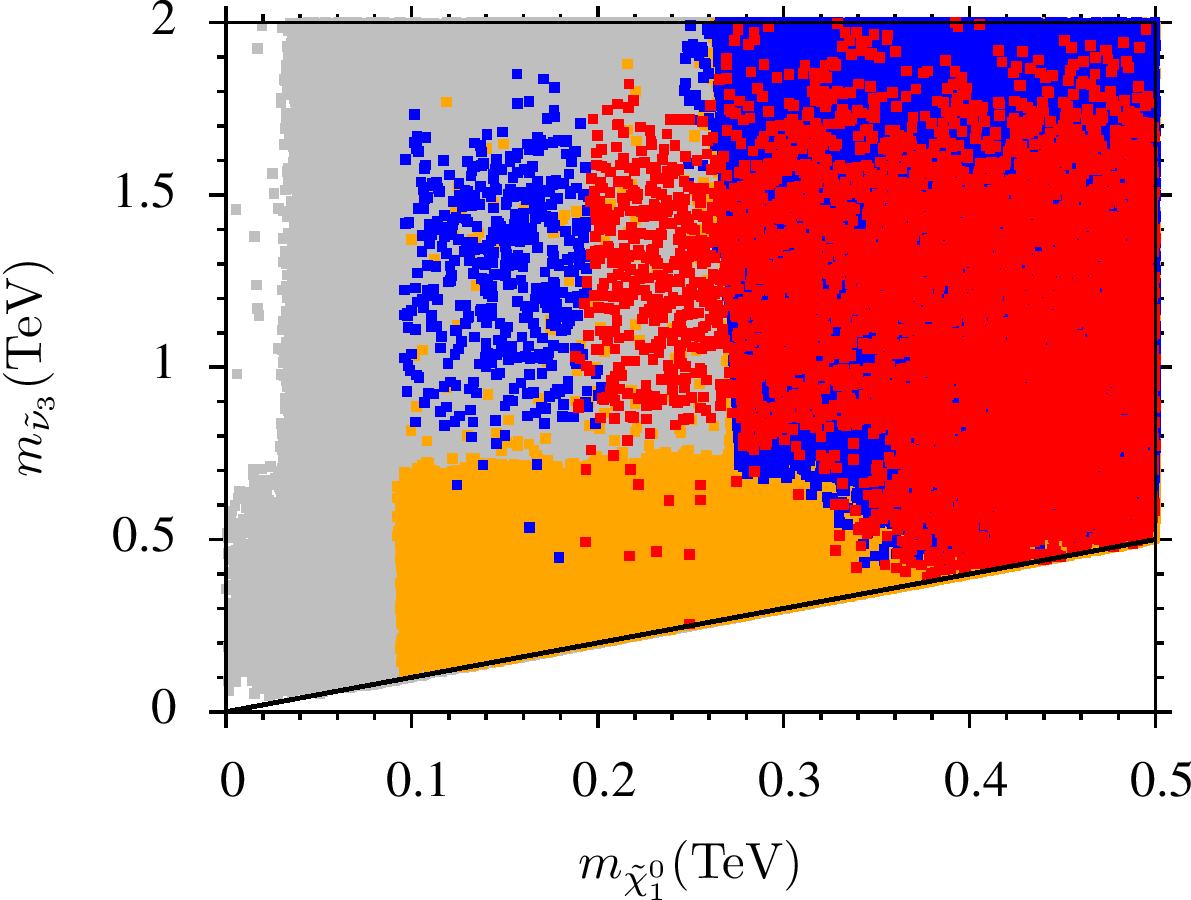}
}

\caption{
Plots in $m_{\tilde \chi_{1}^{0}}-m_{\tilde t_1}$, $m_{\tilde \chi_{1}^{0}}-m_{\tilde \tau_1}$,
$m_{\tilde \chi_{1}^{0}}-m_{A}$,
$m_{\tilde \chi_{1}^{0}}-m_{\tilde \chi_{1}^{\pm}}$ and $m_{\tilde \chi_{1}^{0}}-m_{\tilde \nu_3}$
planes for NUHM2-ISS. The color coding is the same as in Figure~\ref{CMSSM_funda} except that red points
are a subset of blue point solutions and also satisfy bounds for relic  abundance, $0.001 \leq \Omega h^2 \leq 1$.
}
\label{nuhm2_spec}
\end{figure}

The sparticle spectrum for NUHM2-ISS is shown in Figure~\ref{nuhm2_spec}, with color coding the same as in
the previous figures. The top left panel shows an NLSP $\tilde t_1$ in the mass range of
220 - 500~GeV, which can be tested at LHC14. The top panel on right shows  that the NLSP $\tilde \tau_1$ can be as light as 250~GeV, which is somewhat lighter than in
the CMSSM and CMSSM-ISS scenarios.
The bottom left panel shows the presence of
$A$-resonance solutions. This follows from the relatively low $\mu$ values in NUHM2 (Figure. \ref{nuhm2_funda}), and with $m_{H_u}$ and $m_{H_d}$ (or equivalently $\mu$ and $M_A$) being independent parameters.

In the bottom right panel we plot $m_{\tilde \chi_{1}^{\pm}}$ versus $m_{\tilde \chi_{1}^{0}}$ . This graph is very
different from the corresponding one for CMSSM-ISS. In NUHM2-ISS scenario, because of low $\mu$ values, the
chargino can be Higgsino-like, which yields bino-Higgsino mixed
dark matter. This type of solution can be seen along the solid back line. In those cases where $\mu$ is heavy,
the chargino will be wino-like as in the CMSSM-ISS case. Such solutions can are displayed in the second strip
in the graph. We also display a plot in the $m_{\tilde \chi_{1}^{0}}-m_{\tilde \nu_3}$ plane where we show a minimum value $m_{\tilde \nu_3} \approx $ 250~GeV, which is also
consistent with the results reported in Ref.~\cite{Okada:2013ija}.

\begin{figure}[t]
\centering
\subfiguretopcaptrue

\subfigure{
\includegraphics[width=8cm]{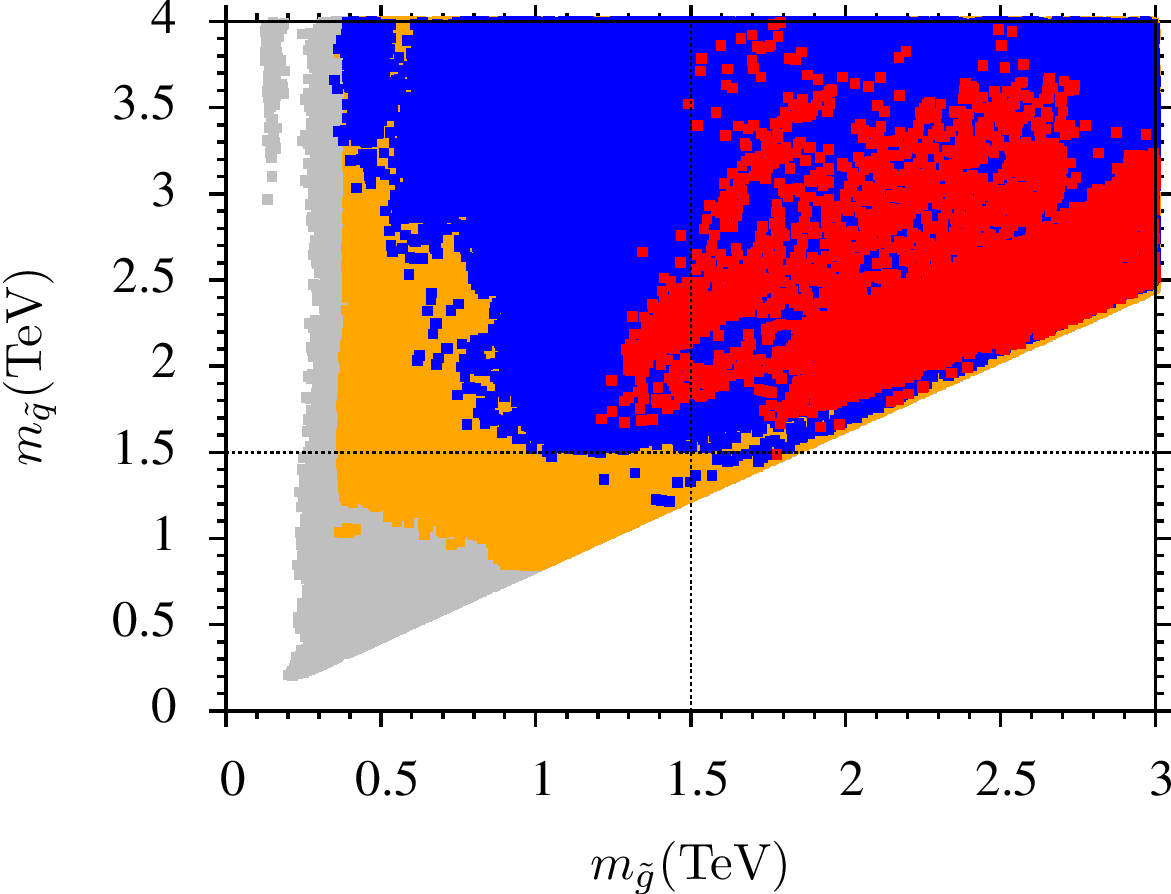}

}
\subfigure{
\includegraphics[width=8cm]{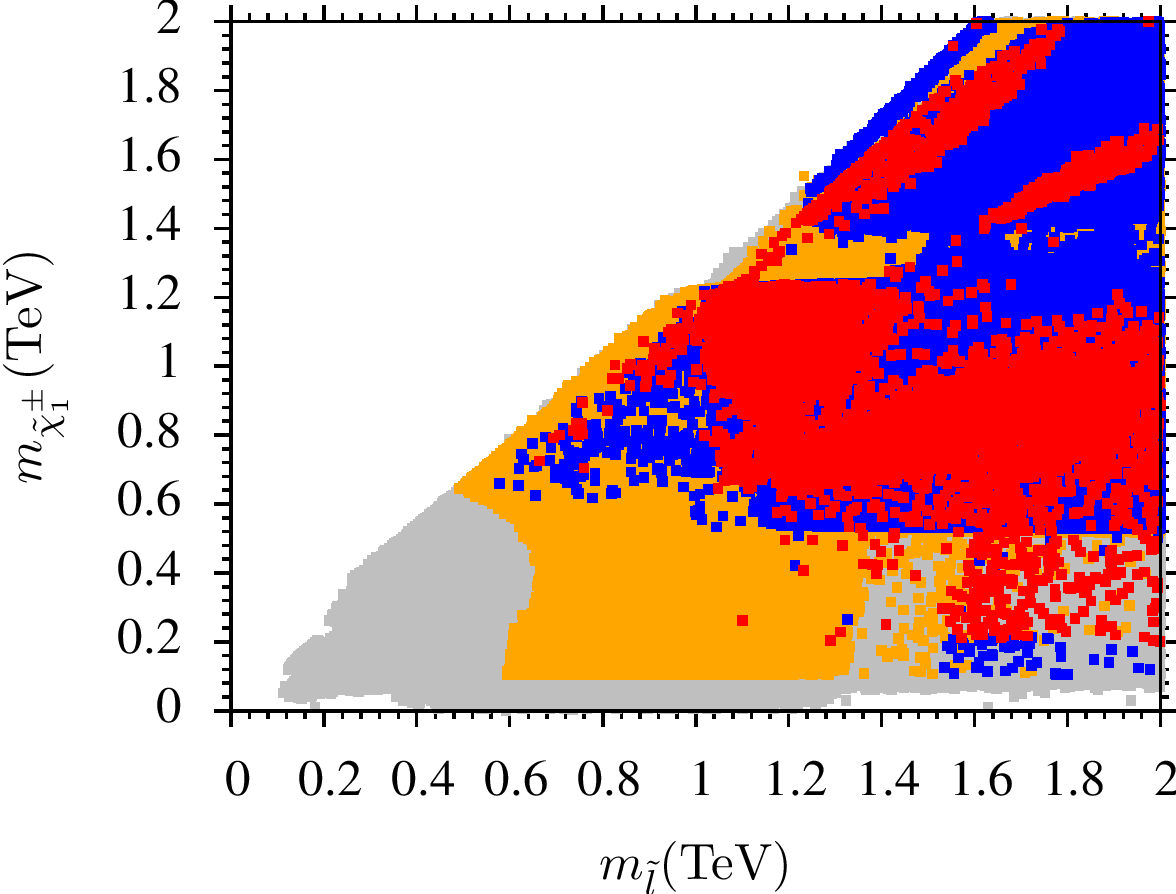}

}

\caption{
Plots in  $m_{\tilde{g}}- m_{\tilde{q}}$ and $m_{\tilde{l}}-m_{\tilde \chi_{1}^{\pm}}$ planes for NUHM2.
In the left panel orange points do not satisfy gluino and first two
 generation squark mass bounds
and red points are a subset of blue point solutions and also satisfy bounds for relic  abundance,
$0.001 \leq \Omega h^2 \leq 1$.
Dashed vertical and horizontal lines stand for current squark and gluino lower mass bounds respectively.
Otherwise color coding is the same as in Figure~\ref{CMSSM_funda}.
}
\label{fig33}

\end{figure}

In Figure~\ref{fig33}
we show graphs in  $m_{\tilde{q}}- m_{\tilde{g}}$ and $m_{\tilde \chi_{1}^{\pm}}-m_{\tilde{l}}$ planes. In the left panel, the orange points do not satisfy the mass bounds for gluinos and
first two generation squarks. The  color coding otherwise is the same as in the previous figures. Dashed vertical and horizontal lines display the current squark and gluino
mass bounds.


\begin{table}[h!]
\centering
\begin{tabular}{lccccc}
\hline
\hline
                 & Point 1  &Point 2 & Point 3 & Point 4 & Point 5 \\
\hline
$m_{0}$         & 2452.3  & 1742.2   & 1573.4   & 1301.9 & 3116 \\
$M_{1/2} $      & 1333.4  & 1292.1   & 968.18   &  1293.1 & 857.6   \\
$A_0/m_0$       & -2.62  & -2.49   & -2.60   &  -2.82 & -2.87 \\
$\tan\beta$     & 53.42  &  12.54    & 26.25   &   22.67  & 18.89\\
$m_{H_d}$       & 4.5484  & 855.55   & 1.8117   &  1.7413 & 737.7\\
$m_{H_u}$       & 2.0939  & 3783.2   & 2661.3   &  3060.1 & 3972\\
\hline
$m_h$          & 125  & 125   & 125  &  125 & 126\\
$m_H$          & 1865 & 1253  & \textbf{882}  & 658  & 2782\\
$m_A$          & 1853 & 1245  & \textbf{876}  & 654  & 2765\\
$m_{H^{\pm}}$  & 1867 & 1256  & 886  & 664 &  2784\\
$\mu$          & 3483 & 6455 & 1448 & 1006 & 3149 \\
\hline
$m_{\tilde{g}}$ & 2971   & 2842  & 2188   & 2816 & 2054       \\
$m_{\tilde{\chi}^0_{1,2}}$
                & \textbf{600}, 1139  & \textbf{556}, 656   & \textbf{423}, 805    & \textbf{563}, 979 & \textbf{388}, 748  \\
$m_{\tilde{\chi}^0_{3,4}}$
               & 3447, 3448   & \textbf{657}, 1080  & 1445, 1450   & 1015, 1103  & 314, 314  \\

$m_{\tilde{\chi}^{\pm}_{1,2}}$
               & 1141, 3448    & 659, 1070  & 807, 1451  & 987, 1097 & 755, 3151 \\
\hline $m_{ \tilde{u}_{L,R}}$
               &3565, 3492   &3052, 3063  &2479, 2483  &2836, 2815 & 3515, 3576  \\
$m_{\tilde{t}_{1,2}}$
               &2195, 2687  &1180, 2302  &1078, 1819   & 1374, 2185 & \textbf{428}, 2303\\
\hline $m_{ \tilde{d}_{L,R}}$
               &3566, 3484    &3053, 2943   &2481, 2406   &2837, 2720 & 3516, 3470    \\
$m_{\tilde{b}_{1,2}}$
               &2628, 2776    & 2305, 2849 &1804, 2153    &2171, 2520 & 2329, 3160   \\
\hline
$m_{\tilde{\nu}_{1}}$
              &2605   & 2021  & 1748   &  1627  &  3225   \\
$m_{\tilde{\nu}_{3}}$ &1503   &804   &818    & \textbf{568} & 1808    \\
\hline
$m_{ \tilde{e}_{L,R}}$
              & 2606, 2502 & 2022, 1611    &1749, 1502  &1630, 1210 & 3222, 3012  \\
$m_{\tilde{\tau}_{1,2}}$
             & \textbf{628}, 1501  & 824, 1536   & 824, 1201 & 588, 972  & 1823, 2693\\
\hline
\hline
$\sigma_{SI}({\rm pb})$
             &1.80$\times 10^{-12}$  & 6.83$\times 10^{-9}$ & 5.11$\times 10^{-11}$ & 5.07$\times 10^{-10}$ & 2.35$\times 10^{-13}$\\
$\sigma_{SD}({\rm pb})$
             & 3.80$\times 10^{-11}$  & 1.00$\times 10^{-5}$ & 2.26$\times 10^{-8}$ & 2.56$\times 10^{-7}$ & 2.1$\times 10^{-10}$\\
$\Omega_{CDM}h^{2}$
            & 0.108  & 0.093    &0.113 & 0.103 & 0.122\\
\hline
\hline
\end{tabular}
\caption{
Masses (in GeV units) and ohter parameters for NUHM2-ISS
 benchmark points satisfying all phenomenological constraints discussed in section \ref{sec:scan}.
Points 1-5 are chosen, respectively, from
 the stau-neutralino coannihilation,
 the bino-Higgsino mixed dark matter,
 the  A-resonance,
 the sneutrino-neutralino coannihilation,
 and the stop-neutralino coannihilation regions.
\label{tab:nuhm2_iss}}
\end{table}


Comparing results from Figures~\ref{fig33} and ~\ref{fig55}, we see very small changes on the lower
mass bounds for the first two generation squarks, and sleptons as well as gluinos, which is what we expected.
But there are many more red points in Figure~\ref{fig33}, because in the NUHM2-ISS case, we have
the additional  $A-$resonance and bino-Higgsino dark  matter solutions for the LSP neutralino relic abundance.
As in the CMSSM-ISS case, we can have squarks and  gluinos in a mass range which can be explored at LHC14.

In Table~\ref{tab:nuhm2_iss} we present five benchmark points for NUHM2-ISS case which satisfy the phenomenological constraints discussed in section~\ref{sec:scan}.
Points 1, 2, 3, 4 and 5 are chosen, respectively, from
 the stau-neutralino coannihilation region, the bino-Higgsino mixed dark matter region, the A-resonance region,
 the sneutrino-neutralino coannihilation region, and the stop-neutralino coannihilation
 region. In all the five benchmark points the first two generation squarks are
 in the mass range 2.4-3.5 TeV, while the first
two generation sleptons lie around 1.6-3~TeV.
Note that for the bino-Higgsino mixed dark matter point the spin independent cross section is $6.83\times 10^{-9}$ pb, which is below the current
XENON100 bounds ~\cite{Xenon100},  but within the reach of XENON1T~\cite{Xenon1T} and
SuperCDMS~\cite{SuperCDMS}.

\section{Conclusions}
\label{conclusions}

The recent discovery at the LHC of a SM-like Higgs boson with mass $m_h\simeq 125$~GeV  puts considerable stress on the MSSM.
With $m_h\lesssim M_Z$ at tree level,
large radiative corrections are required.
Such corrections can be achieved in the MSSM either with multi-TeV stops, or with a large stop trilinear coupling and
stop masses around 1 TeV. In models with universal sfermion masses at $M_{\rm GUT}$, such as CMSSM and NUHM2, this
leads to heavy sleptons and 1st/2nd generation squarks which are near or beyond the ultimate LHC
reach. Various MSSM extensions have been proposed to allow lighter sfermions via additional contributions to the lightest
CP-even Higgs boson mass. In this paper we explored the impact of the inverse seesaw mechanism
on the sparticle mass spectrum.

The ISS mechanism allows an increase of $m_h$ by a few GeV, while simultaneously generating mass for neutrinos via dimension six operators. With a maximal value of the Dirac Yukawa coupling involving the up-type Higgs
doublet, $m_h$ is increased by 2-3~GeV.
As we have shown, this effect allows one to have lighter colored sparticles  in CMSSM
 and NUHM2 scenarios which can be tested at LHC14. For example, in CMSSM-ISS the minimal value of $m_0$ is
 $\sim 400$~GeV, compared to CMSSM where $m_0\gtrsim 800$~GeV.
 Furthermore, requiring neutralino LSP to be the cold dark matter (CDM) pushes $m_0$ to 10-20~TeV range in CMSSM, whereas in CMSSM-ISS
 values as low as $\sim 200$~GeV  are allowed. This means that squarks and gluinos  in CMSSM-ISS lie
 within the reach of LHC14.
Similarly, in NUHM2-ISS squarks and gluinos in 1.5-3~TeV range are consistent with neutralino CDM.
 We have presented several LHC testable benchmark points with the desired neutralino dark matter relic abundance.

\section*{Acknowledgments}
We would like to thank Adeel Ajaib and Cem Salih Un for useful discussions.
This work is supported in part by DOE Grant No. DE-FG02-12ER41808 (I. G., B. H., and Q. S.). A. M. is also supported by a DOE grant.
 This work used the Extreme Science and Engineering Discovery Environment (XSEDE),
 which is supported by the National Science Foundation grant number OCI-1053575. I. G. acknowledges support from the Rustaveli National Science Foundation No. 03/79. B. H. would like to thank the Center for High Energy Physics at Peking University where part of this work was done for hospitality.


\end{document}